\documentclass[sigconf]{acmart}
\usepackage{graphicx}

\usepackage{multicol}

\usepackage{ifthen}
\newboolean{techreport}
\usepackage{caption}
\usepackage{multirow}
\usepackage{subcaption}
\usepackage{hyperref}
\usepackage{amsmath}
\usepackage{tikz}
\usepackage[]{footmisc}
\usepackage{graphicx}% http://ctan.org/pkg/graphicx
\usepackage{pifont}% http://ctan.org/pkg/pifont

\usepackage[ruled,linesnumbered,vlined]{algorithm2e}
\let\oldnl\nl% Store \nl in \oldnl
\newcommand{\nonl}{\renewcommand{\nl}{\let\nl\oldnl}}% Remove line number for one line
\usepackage{tikz}

\definecolor{codegreen}{rgb}{0,0.6,0}
\definecolor{codegray}{rgb}{0.5,0.5,0.5}
\definecolor{codepurple}{HTML}{C42043}
\definecolor{backcolour}{rgb}{1,1,1}

\newcommand\numberstyle[1]{%
    \footnotesize
    \color{codegray}%
    \ttfamily
    \ifnum#1<10 0\fi#1 |%
}

\newtheorem{exmp}{Example}[section]
\newtheorem{theorem}{Theorem}

\newtheorem{definition}{Definition}[section]

\newtheorem{problem}{Problem}

\usepackage[ruled,linesnumbered]{algorithm2e}
\usepackage{algpseudocode}
\SetKw{KwBy}{by}
\SetKwComment{Comment}{/* }{ */}
\RestyleAlgo{ruled}

\usepackage{enumitem}
\renewcommand\footnotetextcopyrightpermission[1]{} % removes footnote with conference information in first column

\begin{document}
\setboolean{techreport}{true}
% ****************** TITLE ****************************************

\title{Display of 3D Illuminations using Flying Light Specks}
\subtitle{Extended Version}
\titlenote{A shorter version in~\cite{shahram2022}, the {\em Proceedings of the 30th ACM International Conference on Multimedia} (MM '22), October 10--14, 2022, Lisboa, Portugal, DOI \url{https://dl.acm.org/doi/10.1145/3503161.3548250}, ISBN 978-1-4503-9203-7/22/10.
See \url{https://github.com/shahramg/FLS-Multimedia2022} for experimental software.}

\author{Shahram Ghandeharizadeh}
\affiliation{%
  \institution{University of Southern California}
    \city{Los Angeles}
    \state{California}
  \country{USA}
}
\email{shahram@usc.edu}

\begin{abstract}
This paper presents techniques to display 3D illuminations using Flying Light Specks, FLSs.  Each FLS is a miniature (hundreds of micrometers) sized drone with one or more light sources to generate different colors and textures with adjustable brightness.  It is network enabled with a processor and local storage.  Synchronized swarms of cooperating FLSs render illumination of virtual objects in a pre-specified 3D volume, an FLS display.  We present techniques to display both static and motion illuminations.  Our display techniques consider the limited flight time of an FLS on a fully charged battery and the duration of time to charge the FLS battery.  Moreover, our techniques assume failure of FLSs is the norm rather than an exception.  We present a hardware and a software architecture for an FLS-display along with a family of techniques to compute flight paths of FLSs for illuminations.  With motion illuminations, one technique (ICF) minimizes the overall distance traveled by the FLSs significantly when compared with the other techniques.
\end{abstract}

\maketitle
\pagestyle{plain} % removes running headers

\section{Introduction}

Unmanned Aerial Vehicles (UAVs) are enabling diverse applications ranging from journalism to entertainment~\cite{Alghamdi2021ArchitectureCA,Shakhatreh2019UnmannedAV,Chung2018ASO}.
A Flying Light Speck, FLS, is a miniature (hundreds of micrometer) sized UAV configured with Red, Green, and Blue light sources to render illuminations~\cite{shahram2021,shahram2022b}.
It is battery powered, network enabled, with some storage, and processing to implement decentralized algorithms.

%\begin{multicols}{2}
%\begin{figure*}
\begin{figure}
%\begin{subfigure}[t]{0.47\columnwidth}
\centering
\includegraphics[width=3.6in]{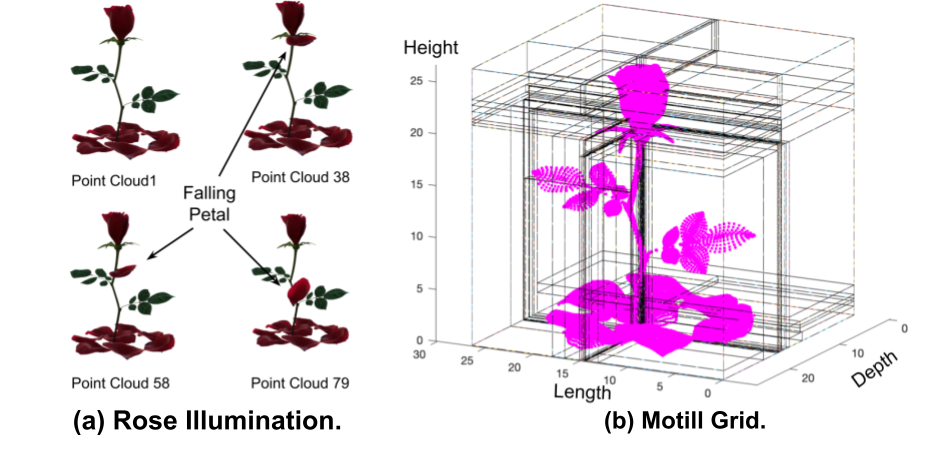}\hfill
%\caption{Rose Illumination.}
%\label{fig:intro:rose}
%\end{subfigure}
%\quad
%\begin{subfigure}[t]{0.47\columnwidth}
%\centering
%\includegraphics[width=\textwidth]{figs/MotillGrid.png}
%\includegraphics[width=1.67in]{figs/MotillGrid.png}
%\caption{Motill Grid.}
%\label{fig:intro:grid}
%\end{subfigure}
\caption{(a) A motion illumination of a rose with a falling petal.  It consists of 115 point clouds rendered at 24 point clouds per second with a 4.79 second display time.  
%%%We show point clouds 1, 38, 58, and 79 with a petal at different stages of falling.  
%Each point cloud requires approximately 65K FLSs.  
1(b) We use Length (L), Height (H), and Depth (D) to identify the dimensions of a 3D coordinate used by both an FLS display and an illumination.  Motill constructs a grid 
%on a motion illumination 
to compute FLS flight paths across a sequence of point clouds.}
\label{fig:intro:rosegrid}
%\end{figure*}
\end{figure}
%\end{multicols}

A swarm of cooperating FLSs are synchronized to render an illumination of a virtual object in a 3D FLS display.
The display is a volume partitioned into a mesh of 3D cells.
Each cell of the display is identified by its length, height, and depth (L,H,D) coordinates, see Figure~\ref{fig:intro:rosegrid}b.  We use the L, H, D coordinate system instead of X, Y, Z because there is no consensus on one definition of the Y and Z axes.  While the picture industry uses the Z axis as the depth, mathematicians use the Y axis for the depth.  It is trivial to map our L, H, D coordinate system to either definition without ambiguity. 

The size of a display cell is dictated by the FLS {\em downwash}, a region of instability caused by the flight of one UAV that adversely impacts other UAVs entering this region~\cite{downwash1,dcad2019,preiss2017,downwash3,Ferrera2018Decentralized3C,planning2019}, e.g., loss of control or unpredictable behavior.
We assume the light emitted by an FLS is larger than a display cell, positioning the FLS at the center of multiple cells along each dimension.

A static illumination is a point cloud.
Each point $p_i$ identifies a 3D coordinate with a value for its color model.
In this study, we assume the RGBA model that specifies the red, green, blue, and alpha color settings.
Hence, a point is a $\{l_i,h_i,d_i,R_i,G_i,B_i,A_i\}$.
%%%While the first three identify the coordinates of the point $p_i$, the last three dictate the intensity of its red, green, and blue lights, respectively.

A motion illumination is a stream.
It may be a stream of $\{p_i,l_i,h_i,$
$d_i,R_i,G_i,B_i,s_i,e_i\}$ where the interval $\{s_i, e_i\}$ specifies when the point identified by $p_i$ should be illuminated relative to the start of the stream.
Alternatively, it maybe a stream of point clouds that must be rendered at a pre-specified rate, e.g., 24 point clouds per second.
Yet another possibility is a hybrid of these two by associating only one $\{s,e\}$ for a point cloud and individual $\{s_i,e_i\}$ for select points.  %%%To prevent interdependence between cells of different point clouds, the $\{s_i,e_i\}$ associated with a point must be shorter than the point cloud's display time.
This paper assumes the second representation, i.e., a sequence of point clouds rendered at a pre-specified rate.
An example is the Rose illumination of Figure~\ref{fig:intro:rosegrid}a with a falling petal.
%It consists of 115 point clouds.  Rendered at 24 point clouds per second, its display time is 4.79 seconds.  
%%%Its physical representation may be a bag file~\cite{} similar to what is used for depthcams.  The bag file supports messages, a simple data structure consisting of typed fields including arrays of primitives, i.e., integer, floating point, etc.

Display of a motion illumination is continuous when FLSs render its $n$ point clouds in a timely manner.
Relative to the illumination of the first point cloud at time $T_1$, each remaining point cloud has a start and an end time stamp relative to $T_1$.  
%%%These time stamps are computed using the rate of point clouds displayed per unit of time, e.g., 24 per second.  
FLSs rendering a point cloud $\Xi_i$ at time $T_j$ must fly to positions dictated by the next point cloud $\Xi_{i+1}$ at 
time $T_{j+\Delta}$. 
$\Delta$ is dictated by the rate of point clouds displayed per unit of time, e.g., $\Delta=\frac{1~Second}{24}$ when 24 point clouds are rendered per second.
%the start time of point cloud $\Xi_{i+1}$ relative to $\Xi_i$.  
Once at their new position, FLSs must render the lighting required by the point cloud $\Xi_{i+1}$.  This process continues until all point clouds of a motion illumination are displayed.

%%%While the position and color of points may change drastically from one scene to another, the changes are anticipated to be minimal within a scene.  This paper focuses on point clouds that constitute a scene.  An example is display of a rose with a falling petal (the Rose clip). This 3D clip consists of 115 point clouds.
%To illustrate, a rose with a falling petal (the Rose clip) consists of 115 point clouds.  Its display time is 4.79 seconds with 24 point clouds displayed per second.  Each point cloud consists of approximately 65K points.  %%%The number of points that change either position or color is approximately 2K from one point cloud to the next.  With one FLS per point, 65K FLSs must take flight with some changing their position and color for 4.79 seconds to render the Rose clip.  

Display of both static and motion illuminations is non-trivial for several reasons.
First, a rendering may require a large number of FLSs.
For example, each point cloud of the Rose illumination consists of 65K points (FLSs).
The Rose illumination is simple.
We anticipate more complex illuminations to consist of millions and potentially billions of points.

Second, FLSs are mechanical devices that fail.  Hence, failures are the norm rather than an exception.  A display requires techniques to render an illumination in the presence of FLSs failing continuously.  
Third, each FLS is battery powered with a fixed flight time.  Its battery requires a certain amount of time to charge.  A key question is what is the relationship between these factors and the extra number of FLSs required to render an illumination?  Sections~\ref{sec:failure} and~\ref{sec:battery} provide an answer.
%%%how these factors translate into the extra number of FLSs required to render an illumination that requires $\alpha$ FLSs.  

Fourth, flight of FLSs may result in collisions.  Computing collision free paths is expensive with tens of UAVs~\cite{collisionfree2012,collisionfree2015,collisionavoidance2018,ReactiveCollisionAvoidance2008,ReactiveCollisionAvoidance2011,ReactiveCollisionAvoidance20112,reactiveColAvoidance2013,downwash1,downwash3,dcad2019,Engelhardt2016FlatnessbasedCF,navigation2017,reactiveColAvMorgan,reactiveColAvBaca,speedAdjust2021,gameCollisionAvoidance2020,gameCollisionAvoidance2017,preiss2017,Ferrera2018Decentralized3C,planning2019,preiss2017whitewash,opticalpositioning1}.  This may be prohibitively expensive with tens of thousands of FLSs.  It may be impractical in the presence of FLSs with limited flight times failing constantly.  Our design philosophy is to detect FLS conflicts when computing flight paths.  We provide this information to the FLSs that participate in the conflict.  When FLSs take flight to render an illumination, they use this information to implement a decentralized technique to avoid collisions.  A simple collision avoidance technique is for the participating FLSs to take turns flying to their destination by using their unique identifier to order themselves.  Such decentralized techniques are implemented using the networking, processing, and storage capabilities of FLSs.

{\bf Contributions} of this paper include:
\begin{itemize}
    \item An architecture for FLS displays to render 3D static and motion illuminations.  (Section~\ref{sec:arch}.)
    \item MinDist and QuotaBalanced algorithms to render a static illumination.  These may be used in either offline or online mode.  In offline mode, they compute a representation that may be stored in a file for future use without re-running the algorithm.  In online mode, they render the illumination without generating files.  Both algorithms are fast and run in tens of milliseconds with illuminations consisting of tens of thousands of points.  (Section~\ref{sec:static}.)
    
    \item Motill, a family of offline algorithms to compute flight paths of FLSs that render 
    %different point clouds of 
    a motion illumination.  One technique, ICF, minimizes the overall distance travelled by FLSs when compared with the other alternatives.  %Its execution time is also faster.
    (Section~\ref{sec:motill}.)
    
    \item A technique that uses standby FLSs to render an illumination in the presence of FLS failures.  With once a month as the mean time to failure of an FLS, the quality of the Rose illumination degrades once every 40 seconds due to failures.  Our proposed techniques enhance this to once a month or more depending on the incurred overhead in the form of additional FLSs. (Section~\ref{sec:failure}.)
    
    \item STAG as a technique that overlaps charging of some FLS batteries with other FLSs rendering an illumination.  We prove optimality of this algorithm in minimizing the total number of FLSs and charging stations for an illumination.  (Section~\ref{sec:battery}.)
    
    \item We open source our software and data pertaining to the Rose illumination for use by the scientific community.  See \url{https://github.com/shahramg/FLS-Multimedia2022} for details. 
 
    %\item Motill as a family of techniques to render a motion illumination.  Motill is highly parallelizable and may use multiple CPU cores to compute FLS flight paths for different portions of an illumination.  
\end{itemize}

The rest of this paper is organized as follows.
Section~\ref{sec:arch} presents a hardware and software architecutre for an FLS display.
Section~\ref{sec:display} presents algorithms to render both static and motion illuminations.
Section~\ref{sec:failure} describes FLS failure handling.
Section~\ref{sec:battery} presents an optimal algorithm for continuously charging battery of FLSs.
Related work is presented in Section~\ref{sec:related}.
We conclude with brief conclusions and future research directions in Section~\ref{sec:future}.

\section{Architecture}\label{sec:arch}

A 3D FLS display consists of a number of software and hardware components.  These include:
\begin{itemize}
    \item Hangars:  FLSs are kept in one or more hangars.  Hangars protect FLSs from external factors that may either damage them or reduce their lifetime.  A hangar may be accessible to one or more Dispatchers, Charging Stations, and Terminus.
    \item Charging Stations:  These charge the battery of FLSs, depositing those with a fully charged battery into a hangar.  A charging station has one or more well defined entry points known to the FLSs.  
    \item Dispatchers:  One or more dispatchers deploy FLSs to render an illumination.  One or more hangars may be accessible to a dispatcher.  Dispatchers may communicate identity, flight path, and deploy time of their FLSs to one another to detect potential FLS crashes.  Dispatchers implement algorithms to avoid potential crashes, see Section~\ref{sec:static}.
    \item Garbage Collectors, GCs:  One or more GCs collect failed FLSs that fall to the bottom of the display and bring them to a Terminus.  A GC may be in the form of a conveyor belt that moves failed FLSs away from the display grid and into entry points that open to the Terminus.
    \item Terminus:  A Terminus has one or more entry points in the display.  In addition to GCs, an FLS that detects it may no longer function properly may fly to a Terminus, see Section~\ref{sec:failure}.  A Terminus may diagnose a failed FLS, perform procedures to recover it to normal mode of operation, and deposit a recovered FLS to a hangar.  
    \item An Orchestrator:  A software component that manages FLSs in Hangars, Charging stations, Dispatchers, Garbage Collectors, and Terminus.  It also manages the storage and network of the Hub.  The Orchestrator may delegate tasks to other components.  For example, it may delegate deployment of FLSs to one or more dispatchers, see Section~\ref{sec:static}.  
    \item Hub:  A Hub provides the processing, storage, and networking capabilities of an FLS display.  It uses an off-the-shelf operating system such as the Linux Ubuntu.  The Hub executes the Orchestrator software that manages and coordinates all aforementioned components.
\end{itemize}
The Orchestrator implements centralized algorithms to render a motion illumination.  It may also implement hybrid centralized and decentralized algorithms that include the participation of the FLSs, dispatchers, or both.  For example, with the parity-based technique of Section~\ref{sec:failure}, the Orchestrator may identify the number of FLSs in a group and their identity.  However, detection of FLS failures and subsequent substitution of a parity FLS for the failed FLS may be performed by the FLSs without the Orchestrator involvement.  At the other end of the spectrum, certain tasks may be implemented in a decentralized manner independent of the Orchestrator.
An example is collision avoidance implemented by FLSs and dispatchers.

\begin{figure}
    \centering
    \includegraphics[width=1.0\linewidth]{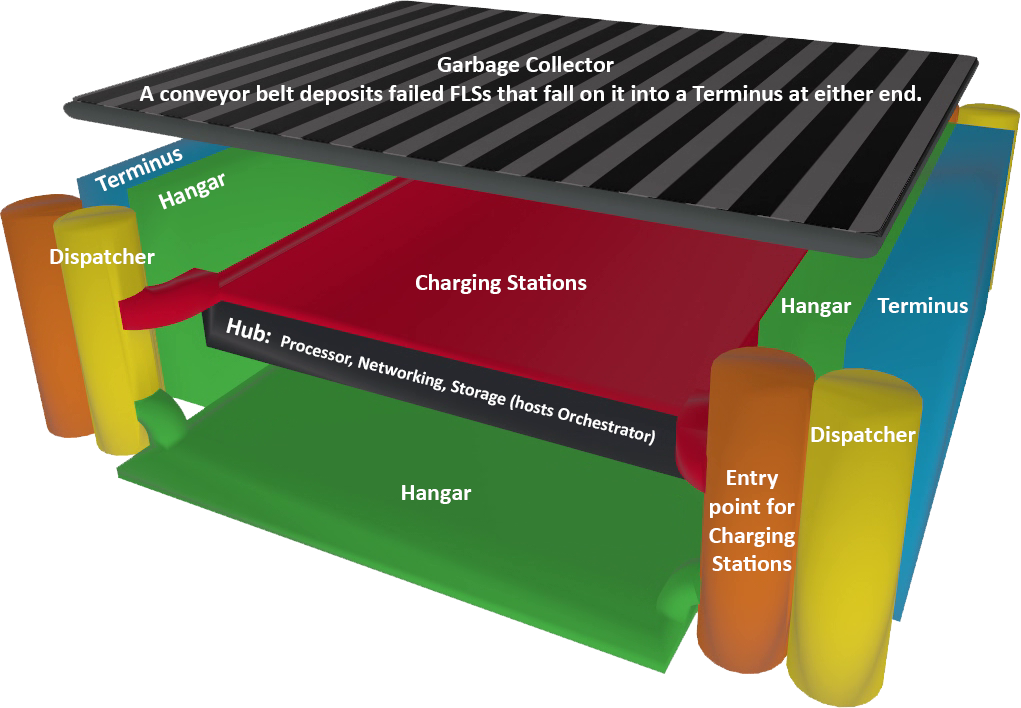}
    \caption{Architecture of an FLS display.}
    \label{fig:arch}
\end{figure}

Figure~\ref{fig:arch} depicts these components as the bottom of an FLS display that sits on a floor or a table top.  The volume that is used to render an illumination is above the garbage collector, i.e., the conveyor belt at the top.  A cylinder at each corner serves as a dispatcher.  Each is accompanied with a cylinder that an FLS flies into to obtain access to the charging stations.  The charging stations deposit fully charged FLSs into hangars located on the two sides and the bottom of the display.  Figure~\ref{fig:arch} shows the Hub as a server blade installed below the charging stations.  Above the charging stations is the garbage collector.  Failed FLSs fall on the garbage collector's conveyor belt that rotates and deposits these FLSs in a Terminus at either end.

Figure~\ref{fig:arch} is one from many possibilities.  With a wall-mounted 3D display and other application use cases, the organization may be completely different.   

\section{Display of 3D Illuminations}\label{sec:display}

To display an illumination, the Orchestrator constructs a 3D mesh 
on the display volume. 
A cell of this mesh is dictated by the downwash of an FLS.  
Assuming an FLS is a quadrotar, a cell may be an 
ellipsoid~\cite{dcad2019,preiss2017,downwash3} or a
cylinders~\cite{Ferrera2018Decentralized3C,planning2019} that 
results in a larger separation along the height dimension.
Each display cell has a unique (L,H,D) coordinate.
It is referenced by one point in the point cloud.

\begin{definition}
A display cell is occupied by one FLS.  Its size is dictated by the downwash of the FLS.  It is identified by a unique (L,H,D) coordinate.
\end{definition}

We use a cuboid to represent an illumination cell rendered by an FLS light source.
The size of this cuboid is dictated by the characteristics of the FLS light source.
It may be either smaller than, 
equal to, or larger than a display cell.
When it is smaller, an FLS may be configured with multiple sets of RGB light sources to render different points.  
When equal to or greater than, an FLS may be configured with one set of RGB light sources. 
When greater than, the FLS's light sources illuminate a point that corresponds to multiple display cells.  
Hence, the FLS is placed at the center of these display cells for illumination.
This scenario is assumed in this paper, enabling an FLS$_i$ to pass by FLS$_j$ as long as it 
stay outside of FLS$_j$'s display cell.
We defer the other two cases to future work.

\begin{definition}
A cuboid consists of six flat faces and eight vertices.  All its faces are rectangles.  A cuboid is a square prism when at least two faces are squares.  A cuboid is a cube when all its six faces are squares.  This paper refers to all as a cuboid.
\end{definition}

\begin{definition}
An illumination cell is a cuboid that is larger than a display cell.
An FLS is positioned such that its rendered light fills one or more faces of the cuboid.  Typically, the illumination cell is occupied by one FLS.  Depending on its size, two or more FLSs may occupy it with at most one rendering its light.  The other FLSs may either be a standby for failure handling (see Section~\ref{sec:failure}) or transitory on their route to their assigned coordinates or a charging station.
\end{definition}

\begin{figure}
    \centering
    \includegraphics[width=1.0\linewidth]{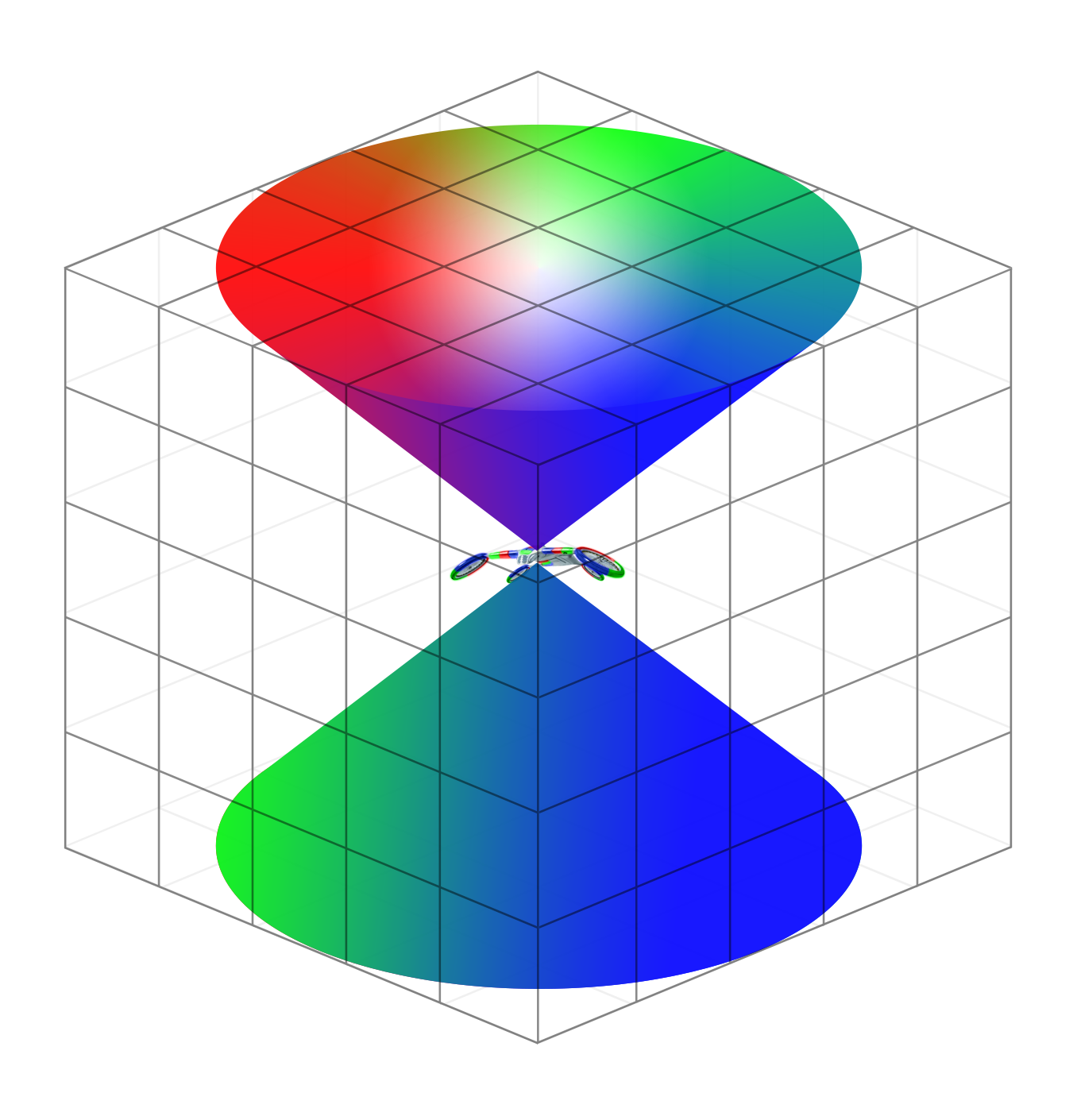}
    \caption{A 5x5x5 illumination cell consisting of 125 display cells with an FLS at its center, occupying the display cell at coordinates 3x3x3 of the illumination cell.  The FLS is rendering RGB lights from its top and bottom to illuminate the top and bottom faces of the illumination cell.}
    \label{fig:illuminationcell}
\end{figure}

Figure~\ref{fig:illuminationcell} shows an illumination cell consisting of L=5, H=5, and D=5 display cells.  It consists of 125 display cells.  It shows an FLS at the center of the illumination cell, at coordinates L=3, H=3, and D=3 of the illumination cells.  The cones show the FLS rendering its top and bottom RGB lights to illuminate the top and bottom faces of the illumination cell.  A maximum of 125 FLSs may occupy this illumination cell with only one rendering its light sources to illuminate one or more faces of the illumination cell.  The other 124 may be transitory on their path to a destination, e.g., another illumination cell, a charging station.  The display cell considers downwash.  Hence, FLSs in an illumination cell will not interfere with one another as long as they occupy a display cell. 

\begin{definition}
A point in a point cloud identifies an illumination cell of an FLS display.  A single FLS renders its light to illuminate the point.
\end{definition}

This section describes display of illuminations assuming FLSs do not fail 
and have unlimited flight times.  
These assumptions are removed in Sections~\ref{sec:failure}
and~\ref{sec:battery}, respectively.

\begin{table}
\small
\centering
\caption{Notations and their definitions.}
\begin{tabular}{cl} \hline
Notation & Definition \\ \hline
$\Xi_i$ & Point cloud $i$. \\ \hline
$\psi$ & Number of dispatchers. \\ \hline
$\theta$ &  Max number of points/FLSs assigned to a cuboid.\\ \hline
$\beta$ & Flight time of an FLS on a fully charged battery. \\ \hline
$\Omega$ & Time to charge an FLS battery fully. \\ \hline
$\kappa_i$ & Number of points assigned to Dispatcher $i$. \\ \hline
$\epsilon_i$ & Flight paths that transition FLSs of $\Xi_i$ to render $\Xi_{i+1}$. \\ \hline
$\alpha$ & Number of points/FLSs in a point cloud.  \\  \hline
$MTTF$ & Mean Time To Failure of an FLS.  \\  \hline
\end{tabular}
\label{tab:term}
\end{table}

\subsection{Display of Static Illumination}\label{sec:static}
An algorithm may illuminate a point cloud based on different objectives.  
An example objective is to minimize the total distance traveled by FLSs to arrive at the coordinate of their assigned point, i.e., an illumination cell.
The formal definition of this assignment problem is as follows.

\begin{problem}
An illumination consists of $\alpha$ points and a display consists of $\psi$ dispatchers.
The distance from a dispatcher to a point is fixed, Distance($Dispatch_i$, $P_j$).
%There are $\alpha$ points and $\psi$ dispatchers with a fixed distance from a dispatcher to a point, Distance($Dispatch_i$, $P_j$).  
Assign each point to a dispatcher such that one and only one point is assigned to a dispatcher and 
the total distance for a dispatcher and its assigned point is minimized,
%the total distance of the assignments is minimal,
i.e., minimize $\sum_{i=1}^{\psi}\sum_{j=1}^{\kappa_i}{Distance(Dispatch_i, P_{j}})$ 
where $\kappa_i$ is the number of points assigned to a dispatcher $i$.
\end{problem}

MinDist is an algorithm that iterates each point, computes the distance of the point to a dispatcher, and assigns the point to the dispatcher with the shortest euclidean distance.
See the first for loop of MinDist, Lines 1-10 of Alg~\ref{alg:mindist}.

Subsequently, each dispatcher sorts its assigned points in descending order based on their distance from its location.  It deploys FLSs to render points starting with the farthest away one first, see the second for loop of MinDist, Lines 11-14.  This minimizes the possibility of dispatched FLSs from colliding with one another. 

The first for loop of MinDist, Lines 1-10, is sequential and may be implemented by the Orchestrator.  The second for loop may be processed by each dispatcher in parallel and independent of the Orchestrator.
%This is realized by the Orchestrator delegating the sorting of points and deployment of their respective FLSs to a dispatcher for its assigned points.

The complexity of the first loop is O($\psi \times \alpha$) where $\alpha$ is the number of points in a point cloud and $\psi$ is the number of dispatchers.
The complexity of the second loop is dictated by the number of points $\kappa_i$ assigned to a dispatcher $i$ and its algorithm to sort the points based on their distance, e.g., with the QuickSort algorithm, the complexity is O($\kappa_i^2$) for dispatcher $i$.

\begin{algorithm}
\caption{MinDist
%A point assignment and deployment heuristic to minimize travel distance.  The second for loop may be processed by each dispatcher in parallel with other dispatchers.
}\label{alg:mindist}
\begin{small}
  \For{$i\gets1$ \KwTo $\alpha$ \KwBy $1$\Comment*[r]{Iterate $\alpha$ points}}{
    $min \gets MaxInteger$\;
    \For{$j \gets1$ \KwTo $\psi$ \KwBy $1$\Comment*[r]{Iterate $\psi$ dispatchers}}{
        $\Delta \gets$ Distance ($Point_i$, $Dispatch_j$)\;
        \If{$\Delta < min$}
        {
        $min \gets \Delta$\;
        $tgt \gets j$\;
        }
    }
    $Dispatch_{tgt} \gets Point_i$\;
    }
    \For{$j \gets1$ \KwTo $\psi$ \KwBy $1$\Comment*[r]{Deploy FLSs}}{
    $Dispatch_j$ sorts its points in descending distance order\;
    $Dispatch_j$ deploys points starting with the farthest away one\;
    }
    \end{small}
\end{algorithm}

MinDist has several limitations.  First, it may result in a slow rendering of an illumination by utilizing a subset of dispatchers more often than others.  This happens when most of the points in a cloud are clustered in close proximity of a few dispatchers.
While these dispatchers deploy most of the FLSs sequentially, other dispatchers sit idle.
See discussions of Table~\ref{tbl:staticCMP} in Section~\ref{sec:staticeval}.

Second, MinDist assumes a dispatcher may access all hangars and their FLSs.  This assumption is violated when hangars are physically partitioned across dispatchers.  MinDist may not render an illumination that consists of a cluster of points in close proximity of a dispatcher with insufficient number of FLSs.

We now present Alg~\ref{alg:balance}, QuotaBlanced, that considers the distance travelled by FLSs, FLS speed, the number of dispatchers, the number of FLSs accessible to a dispatcher, and the rate at which a dispatcher may deploy FLSs.
It assigns a quota to each dispatcher that is reduced as a function of the travel time by its deployed FLSs.  The idea is to have a dispatcher that is very far from the points of a cloud to deploy some FLSs but not as many FLSs as those dispatchers that are in close proximity to the points of the cloud. 

\begin{algorithm}
\caption{QuotaBalanced
%A point assignment to minimize time required to render an illumination.
}\label{alg:balance}
\begin{small}
  $f \gets FLS~deployment~rate, \frac{FLSs}{second}$\;
  $S \gets FLS~Speed$\;
  $\varpi[1..\psi] \gets \frac{\alpha}{\psi \times f}$\Comment*[r]{Quota for each dispatcher}
  $\{Active\} \gets Dispatchers~with~FLSs~and~quota~>~0$\;
  \For{$i\gets1$ \KwTo $\alpha$ \KwBy $1$}{
  %\Comment{Lines~1-13~of~Alg~\ref{alg:mindist} compute $Dispatch_{tgt}$.}
    $Dispatch_{tgt} \gets$ The active dispatcher closest to $Point_i$\;
    $\Delta \gets$ Distance ($Dispatch_{tgt}$,$Point_i$)\;
    $Dispatch_{tgt} \gets Point_i$\;
    \If{$Dispatch_{tgt}$~has~zero~FLSs}{
        $\{Active\}=\{Active\}-Dispatch_{tgt}$\;
        Remove~$Dispatch_{tgt}$~from~further~consideration\;
    }
    $t \gets \frac{\Delta}{S}$ \Comment*[r]{FLS travel time}
    $\varpi[Dispatch_{tgt}]= \varpi[Dispatch_{tgt}] - t$\;
    \If{$\varpi[Dispatch_{tgt}] \leq 0$}{
        $\{Active\}=\{Active\}-Dispatch_{tgt}$\;
    }
    \If{$\{Active\} == \varnothing$}{
        $\varpi[1..\psi] \gets \frac{\alpha-i}{\psi \times f}$\Comment*[r]{Re-compute dispatcher quota}
         $\{Active\} \gets$ Dispatcher$_i$~with~FLS[i]>0~and~$\varpi[i]>0$\;
    }
    }

    \For{$j \gets1$ \KwTo $\psi$ \KwBy $1$}{
    $Dispatch_j$ sorts its points in descending distance order\;
    $Dispatch_j$ deploys points starting with the farthest away one\;
    }
\end{small}
\end{algorithm}

Distance is an approximation of travel time.  The time for an FLS to fly from a dispatcher to its display cell is a function of the FLS speed.
A dispatcher may be far from the point cloud.  However, if FLSs are extremely fast then their travel time may become insignificant to the time to deploy FLSs.
This motivates an algorithm that requires each dispatcher to deploy its fair share\footnote{This changes the problem definition.  We are no longer minimizing total distance travelled.} of FLSs while considering travel time of FLSs.  QuotaBalanced is one such algorithm. 

QuotaBalanced assumes each dispatcher may deploy $f$ FLSs per time unit (Line 1 of Alg~\ref{alg:balance}) and dispatcher $i$ has access to a fixed number of FLSs.  The granularity of its quota is time units required for a dispatcher to deploy its fair share of FLSs for the point cloud, $\frac{\alpha}{\psi \times f}$.  This algorithm converts distance to time using the speed of an FLS, see Line 12.

%%QuotaBalanced continues to use minimum distance to assign points to a dispatcher.  

In each iteration, QuotaBalanced reduces the quota of a dispatcher by the FLS travel time to its assigned point.  Once the quota of a disptacher is exhausted, it is removed from the list of active dispatchers.  This causes QuotaBalanced to assign points to other dispatchers that are not necessarily as close.  However, the quota of these dispatchers are reduced by a larger value because they are farther away, i.e., time to travel is longer.  Hence, these dispatchers will be removed from the active list after a fewer point assignments.

The quota of all dispatchers may be exhausted while some points remain unassigned.
QuotaBalanced re-computes the quota of each dispatcher using the remaining points, see Lines 16-18.  It continues to assigns points to the dispatchers until their quotas are exhausted.  This process repeats until all points are assigned to dispatchers.
The number of repetitions is 1393
with the point cloud of Section~\ref{sec:staticeval}.  

A dispatcher with no FLSs is permanently removed from the Active list (Lines 9-11).
This causes other dispatchers to deploy FLSs to render the illumination. 

The worst case complexity of this heuristic is O($\psi \times \alpha$).  It may perform fewer comparisons than MinDist because dispatchers are removed from the list.  At the same time, in each iteration, it performs more work when compared with MinDist because it must consider the travel time of an FLS to adjust the quota of a dispatcher and potentially reset the quota of all dispatchers.
In our experiments of Section~\ref{sec:staticeval}, MinDist and QuotaBalanced provided comparable execution times. 

Once points are assigned to dispatchers, a dispatcher may deploy FLSs similar to the discussion of MinDist, see Lines 19-21.  
%Similar to MinDist, its second for loop deploys FLSs.  This step may be performed by dispatchers in parallel similar to MinDist. 
There is one difference.  With QuotaBalanced, dispatchers may deploy FLSs that cross paths and potentially crash with one another.  Dispatchers may share the flight path and deployment time of their FLSs with one another to detect such potential crashes.  Prior to deploying an FLS, a dispatcher may detect a conflict with other deployed FLSs traveling to their target point.  It may implement a variety of techniques to prevent crashes.  
A simple technique is to delay the deployment of the FLS by the flight time of the conflicting FLS, increasing the latency to render an illumination. 
Alternatively, the dispatcher may compute a different flight path for its FLS, eliminating a possible crash.

%%%A simple preventive technique will require the detecting dispatcher to delay deployment of its FLS by the flight time of the conflicting FLS, increasing the latency to render an illumination.  A more sophisticated technique may require the detecting dispatcher to compute a different flight path for its FLS, eliminating a possible crashes.  

\subsubsection{A Comparison}\label{sec:staticeval}

\begin{figure}[!ht]
    \centering
    \includegraphics[width=1.0\linewidth]{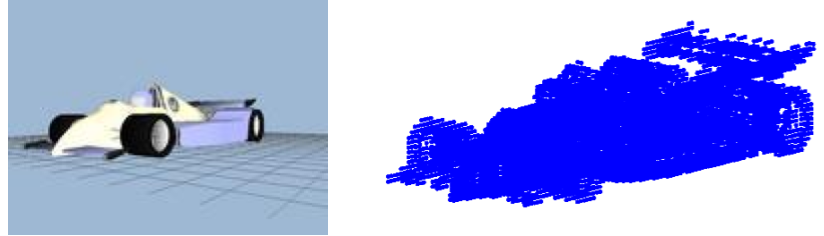}
%\caption{Thumbnail of the point cloud m1510.}
\caption{Shape m1510, thumbnail and its point cloud.}
    \label{fig:m1510}
\end{figure}

We use the Princeton Shape Benchmark~\cite{princetonbenchmark} to highlight the quantitative and qualitative differences between MinDist and QuotaBalanced.  Its database consists of 1,814 3D models.  We present results from 1 model, the race car (m1510), as the findings are identical across all models.  Figure~\ref{fig:m1510} shows the thumbnail for this model and its derived point cloud consisting of 11,894 points. 

We simulate an FLS display that is a cube with length=100, height=100, and depth=100 cells.  This display consists of 8 dispatchers, one at each corner of the cuboid.  Each dispatcher may deploy 10 FLSs per second, one FLS every 100 milliseconds.  The speed of each FLS is 4 cells per second.  An FLS flies along a straight line from the location of its dispatcher to the coordinates of its assigned point.

While QuotaBalanced uses all 8 dispatchers\footnote{The difference between the dispatcher that deploys the most number of FLSs and the one that deploys the fewest number of FLSs is 197, approximately 20\%.} to deploy FLSs, MinDist uses only 2 at the bottom corner\footnote{One deploys 6613 and the other deploys 5281 FLSs.} of the display.  Hence, QuotaBalanced enhances latency four folds, see 1st row of Table~\ref{tbl:staticCMP}.
This comes at a cost, namely, an increase in the total distance travelled by the FLSs.
This metric is more than 2x higher with
QuotaBalanced, see 2nd row of Table~\ref{tbl:staticCMP}.
This translates into a higher energy consumption to enhance latency.

QuotaBalanced exhausts the list of active dispatchers and resets their quota (Lines 16-18 of Alg~\ref{alg:balance}) 1393 times with this model.

Our simulated dispatchers detect when the path of their deployed FLSs intersect one another, identifying a conflict that may result in FLSs crashes.  
QuotaBalanced incurs 35 such conflicts.
Not every detected conflict is a crash because one FLS may fly past the crash point in advance of the other conflicting FLSs.  In our study, FLSs must be in 20\% proximity of one another to be considered as conflicting.
Table~\ref{tbl:staticCMP} shows QuotaBalanced incurs 12 such conflicts.  This is because dispatchers deploy FLSs that are farther away from a point in the display coordinate system than the other dispatchers.  This is a small percentage (0.1\%) of the 11,894 deployed FLSs by the 8 dispatchers.  See the last paragraph of the previous section on how to eliminate possible crashes.

%When a dispatcher deploys an FLS $i$, our simulator detects which of the previously deployed FLSs fall on its trajectory.  Each detected FLS $j$ is counted as a conflict with the potential to cause FLS crash.  The simulator resolves these conflicts by requiring (a) FLS $j$ to assume the identity of FLS $i$ and fly to the coordinates of the point illuminated by FLS $i$, and (b) FLS $i$ to assume the lighting responsibility of FLS $j$ and stop at the coordinates of FLS $j$.  The number of such conflicts are shown in the last row of Table~\ref{tbl:staticCMP}.  MinDist has zero conflicts because only two dispatchers deploy FLSs, each starting with its farthest away point (that is non-overlapping with the points of the other dispatcher).  QuotaBalanced observes a higher number of conflicts because its dispatchers deploy FLSs that cross each other's paths.    

\begin{table}
\begin{small}
  \caption{MinDist vs. QuotaBalanced}
  \label{tbl:staticCMP}
  \begin{tabular}{c|cc}
    \toprule
    & MinDist & QuotaBalanced\\
    \midrule
    Illumination Latency (Seconds) & 661 & 163 \\
    Distance Travelled (Cells) & 494,938 & 1,122,947\\
    Intersecting flight paths & 0 & 35 \\
    FLS Conflicts & 0 & 12 \\
    Execution Time (Milliseconds) & 27.71 & 27.44 \\
    %$\Psi^2_1$ & 1 in 40,000& Unexplained usage\\
  \bottomrule
\end{tabular}
\end{small}
\end{table}

\subsection{Display of Motion Illuminations}
We assume a motion illumination consists of a sequence of point clouds displayed at a pre-specified rate.  
See 
%the Rose illumination of 
Figure~\ref{fig:intro:rosegrid}a.

%We focus on intra-scene point clouds with minimal changes from one point cloud to the next.
%For example, the Rose illumination of Figure~\ref{fig:intro:rose} with a falling petal consists of 115 point clouds.  Its display time is 4.79 seconds with 24 point clouds displayed per second. Each point cloud consists of approximately 65K points.

Assuming an FLS corresponds to a point, a display must compute both travel path of FLSs and their change of color from one point cloud $\Xi_i$ to the next point cloud $\Xi_{i+1}$.  While these changes may be minor with point clouds that constitute a scene, they may be drastic from the last point cloud of one scene to the first point cloud of its following scene.  This paper focuses on computing the intra-scene travel paths, deferring inter-scene travel paths to future work. 

To render a scene, the display must assign an FLS to each point of its first point cloud $\Xi_1$.  This is identical to rendering a static illumination.
Thus, either MinDist or QuotaBalanced maybe used.
To render its subsequent point cloud $\Xi_2$, the Orchestrator must compute whether:
\begin{enumerate}
    \item\label{simple:scenario1} $\Xi_2$ consists of more points than $\Xi_1$, requiring additional FLSs to render it.
    In general, dark FLSs from a previous point cloud (say $\Xi_{i-1}$) may be used to render $\Xi_{i+1}$.
    If none are available then FLSs are deployed by a dispatcher.
    \item\label{simple:fewer} $\Xi_2$ consists of fewer points than $\Xi_1$, requiring some FLSs illuminating $\Xi_1$ to either go dark or fly to a charging station.  Dark FLSs may be used in a subsequent point cloud, say $\Xi_3$.  This may minimize the overall distance travelled by FLSs.  This is because requiring FLSs to fly back to a charging station for $\Xi_2$ only to dispatch FLSs to illuminate $\Xi_3$ may result in a longer total travel distance. 
    \item\label{simple:colorchangeonly} An FLS illuminating $\Xi_1$ remains stationary and changes color in $\Xi_2$.
    \item\label{simple:flyoldnewcolor} An FLS illuminating $\Xi_1$ flies to a new point identified by $\Xi_2$ and displays either the same or a different color.
    %\item\label{simple:flynewcolor} A $\Xi_1$ FLS flies to a new position and displays a different color in $\Xi_2$.
    \item\label{simple:nochange} An FLS illuminating $\Xi_1$ remains stationary and continues to display its current color in $\Xi_2$.  This is the scenario where the point in $\Xi_1$ and $\Xi_2$ are identical, i.e., identify the same illumination cell and render the same color.
\end{enumerate}
Any and all combinations of these possibilities may apply when considering two point clouds $\Xi_i$ and $\Xi_{i+1}$.
%%%A point cloud of the Rose illumination consists of 65K points.  All the above possibilities may be present when considering two point clouds $\Xi_i$ and $\Xi_{i+1}$ of this clip.

This section presents two offline algorithms, Simple and Motill, to detect the alternative scenarios and compute FLS flight paths and color renderings.
Similar to the MPEG encoding technique, a system may execute these algorithms once for a motion illumination, store their computed flight paths, and reuse this information when rendering a motion illumination repeatedly either for the same user or different users.  

\subsubsection{Simple}\label{sec:simple}
Simple is an offline algorithm that consists of two steps.
%Step 1 computes the flight path of FLSs between all pairs of point clouds $\Xi_i$ and $\Xi_{i+1}$ that constitute a scene.  This step computes a set of flight paths $\{\epsilon_i\}$ for point cloud $\Xi_i$.  These flight paths navigate a subset (potentially all) FLSs that are rendering $\Xi_i$ to illuminate $\Xi_{i+1}$.  %Step 1 identifies those FLSs of $\Xi_i$ that are extras.  We use set $\{\delta_i\}$ to denote these FLSs.  
%Step 1 also identifies those FLSs of $\Xi_i$ that are extras, $\{\delta_i\}$, and those points of $\Xi_{i+1}$ that have no assigned FLSs, $\{\mu_{i+1}\}$. 
Step 1 computes flight paths that transition one point cloud to the next.  It also identifies FLSs that are extras from one point cloud to next.  And, points that may require FLSs to be deployed by a dispatchers.
More formally, with $n$ point clouds, $\{\Xi_1, \Xi_2, \cdots, \Xi_n\}$, Step 1 computes four sets.
Flight paths for the first $n-1$ point clouds, denoted $\{\epsilon_1, \epsilon_2, \cdots, \epsilon_{n-1}\}$.
Change of color for the FLSs used in the first $n-1$ point clouds, denoted $\{\gamma_1, \gamma_2, \cdots, \gamma_{n-1}\}$.
Extra FLSs for the first $n-1$ point clouds, denoted $\{\delta_1, \delta_2, \cdots, \delta_{n-1}\}$.
There are extra FLSs when $\Xi_i$ consists of more FLSs than $\Xi_{i+1}$.
Points with no assigned FLSs for the last $n-1$ point clouds, denoted $\{\mu_2, \mu_3, \cdots, \mu_{n}\}$.
$\Xi_i$ may have points with no assigned FLSs when it consists of more points than $\Xi_{i-1}$.

In Step 2, Simple processes $\{\delta_i\}$ to decide whether one or more of its FLSs should stay in the display grid and go dark or fly back to a charging station.  The dark FLSs are assigned to a point identified by a $\{\mu_{j>i}\}$.  Step 2 may schedule a dispatcher to deploy FLSs to illuminate points of $\{\mu_j\}$.

Step 2 is required when there are extra FLSs, points with no assigned FLSs, or both, i.e., either $\{\delta_i\}$, $\{\mu_j\}$, or both are not empty.
Step 1 produces empty $\{\delta_i\}$ and $\{\mu_j\}$ when the different point clouds consist of the same number of points.  In this case, Simple does not execute Step 2.

Simple may optimize for total distance travelled, the amount of energy used, the time required to execute flight paths, or a hybrid of these.  The hybrid may assign weights to different criterion.  This paper focuses on minimizing the total distance travelled by FLSs, deferring other possibilities to future work.  Below, we detail the two steps in turn.

%%Simple processes two point clouds $\Xi_i$ and $\Xi_{i+1}$ to detect and resolve the different scenarios.
%%%It resolves Scenarios~\ref{simple:scenario1}-\ref{simple:flyoldnewcolor} by computing flight paths for some FLSs belonging to $\Xi_i$ to illuminate $\Xi_{i+1}$. 
%%Its objective is to minimize the total distance travelled by the FLSs.
%%This includes distance travelled by FLSs that return to a charging station or deployed by a dispatcher.
%flights by FLSs deployed by a dispatcher or returned to a charging station.

%%\subsubsection{Detection of different scenarios} 
In Step 1, Simple constructs a hash table on the coordinates of $\Xi_i$ and probes it with coordinates of $\Xi_{i+1}$.  For each match, it checks whether the color of the probing point is the same or different.  If they are the same, it has identified a point that belongs to Scenario~\ref{simple:nochange}.
If different then it has identified a point that belongs to Scenario~\ref{simple:colorchangeonly}.
It adds this change to $\gamma_i$.

If the point from $\Xi_{i+1}$ has no match then it is added to the set $\{\mu_{i+1}\}$.  This point is a candidate for Scenario~\ref{simple:flyoldnewcolor} if $\Xi_{i}$ and $\Xi_{i+1}$ consist of the same number of points.
It is a candidate for Scenario~\ref{simple:scenario1} (Scenario~\ref{simple:fewer})
if $\Xi_{i}$ has more (fewer) points than $\Xi_{i+1}$.

Step 1 constructs $\{\delta_i\}$ as follows.  It deletes a hash table entry that matches a coordinate of $\Xi_{i+1}$.
Once all points of $\Xi_{i+1}$ have probed the hash table, 
Simple enumerates those points that remain in the hash table and assigns them to $\{\delta_i\}$.
FLSs in $\{\delta_i\}$ correspond to $\Xi_i$ points with no matches in $\Xi_{i+1}$.
%They are used to render $\Xi_i$ and are freed in $\Xi_{i+1}$.
%Simple may require these to fly to a new coordinate and change color (includes going dark) to render a point of $\Xi_{i+1}$ that appears in $\{\mu_{i+1}\}$, addressing Scenarios~\ref{simple:fewer} and~\ref{simple:flyoldnewcolor}.
%With those that address Scenario~\ref{simple:flyoldnewcolor}, some may go dark to address the same scenario for a subsequent point cloud, say $\Xi_{i+2}$.  This is handled in this step as detailed below.

%that identifies those FLSs that change their position from point cloud $C_i$ to the next point cloud $C_{i+1}$ and computes their flight paths.  Its objective is to minimize the total distance travelled by FLSs.

%Given two point clouds that follow one another in a scene, $C_i$ and $C_{i+1}$, a simple technique may compute those points whose position does not change from $C_i$ to $C_{i+1}$.  Next, it identifies points present $C_i$ and missing from $C_{i+1}$, $\delta_i$.  And, identify points present in $C_{i+1}$ and missing from $C_i$, $\delta_{i+1}$.  If the number of points in $\delta_i$ is greater than $\delta_{i+1}$ then some FLSs must either go dark or fly back to a dispatcher.  If the number of points $\delta_i$ is less than $\delta_{i+1}$ then either dark in-flight FLSs of $C_i$ must start to render color or dispatchers must deploy FLSs to render missing points of $C_{i+1}$.  

Step 1 computes flight paths by mapping points of $\{\delta_i\}$ to $\{\mu_{i+1}\}$.
It minimizes the overall distance using the following greedy heuristic.
It computes the distance\footnote{Distance may be replaced with the amount of energy required or flight time to consider different objectives.} between every pairing of a $\{\delta_i\}$ point with a $\{\mu_{i+1}\}$ point.  
It sorts these pairing in ascending distance.
It selects the first pairing, ($P_{1,\delta},P_{1,\mu}$), and assigns the FLS at the display cell $P_{1,\delta}$ to fly to the vacant display cell identifies by $P_{1,\mu}$ that requires illumination.
This is a flight path with a source coordinate $P_{1,\delta}$ and a destination coordinate $P_{1,\mu}$ that transitions $\Xi_i$ to render $\Xi_{i+1}$.
It is added to the set of flight paths $\{\epsilon_i\}$.
%It adds this flight path to its list of FLS fly paths for $\Xi_{i}$.  
Simple removes the coordinates $P_{1,\delta}$ and $P_{1,\mu}$ from its sorted list of possibilities, and 
from $\{\delta_i\}$ and $\{\mu_{i+1}\}$, respectively.
Step 1 continues with the next pairing, repeating this process until either $\{\delta_i\}$, $\{\mu_{i+1}\}$, or both are empty.

If both $\{\delta_i\}$ and $\{\mu_{j}\}$ are empty then Simple terminates.  Otherwise, it proceeds to Step 2.

Step 2 processes $\{\delta_i\}$ and $\{\mu_{j}\}$ sets.
If $\{\delta_i\}$ is not empty and $\{\mu_{j}\}$ is empty then all FLSs in $\{\delta_i\}$ are required to fly back to a charging station after rendering their point cloud $\Xi_i$.
If $\{\mu_j\}$ is not empty and $\{\delta_i\}$ is empty then dispatchers are scheduled to deploy FLSs to points identified by $\{\mu_j\}$ to illuminate point cloud $\Xi_j$.
With both scenarios, Step 2 is complete and Simple terminates.
%Below, we detail Step 2 when both $\{\delta_i\}$ and $\{\mu_j\}$ are not empty.

When both $\{\delta_i\}$ and $\{\mu_{j}\}$ are not empty then Step 2 processes the two sets as follows.
If there is a $\{\delta_i\}$ with no following $\{\mu_{j}\}$ ($i < j$) then all FLSs identified by $\{\delta_i\}$ are scheduled to fly back to a charging station.  Otherwise, Step 2 computes the distance\footnote{Distance may be replaced with the amount of energy required or flight time to consider different objectives.} between every pairing of a $\{\delta_i\}$ point with a $\{\mu_{j}\}$ point.
It sorts these pairing in ascending distance.
It selects the first pairing, ($P_{1,\delta},P_{1,\mu}$) with distance $\tau_1$, and computes the total distance travelled if $P_{1,\delta}$ flies to a charging station and an FLS is deployed by a dispatcher closest to $P_{1,\mu}$ to illuminate $P_{1,\mu}$.  If the total distance is smaller than $\tau_1$ then the FLS identified by $P_{1,\delta}$ is scheduled to fly back to a charging station\footnote{In this scenario, Step 2 may consider other following $\mu_{k>j}$.  This may identify a destination point in a point cloud $\Xi_k$ for the FLS in $\delta_i$, producing a flight path stored in $\epsilon_i$.}.
Otherwise, Step 2 computes the flight path from $P_{1,\delta}$ to $P_{1,\mu}$ and adds this path to $\epsilon_i$.
Step 2 removes $P_{1,\delta}$ and $P_{1,\mu}$ from its sorted table, and from $\delta_i$ and $\{\mu_{j}\}$, respectively.

Step 2 processes other special cases not detailed here.  For example, if there is a $\{\mu_{j}\}$ with no preceding $\{\delta_i\}$ ($i<j$) then Step 2 schedules FLSs to be deployed by a dispatcher to illuminate points identified by $\{\mu_{j}\}$.

\subsubsection{An Analysis}\label{sec:simpleanalysis}
In our experiments with the Rose illumination, Step 1 requires approximately 600 seconds to compute flight paths for FLSs that constitute two consecutive point clouds.  
This step wastes time and resources by computing a pairing of each FLS in $\{\mu_{i+1}\}$ with each FLS in $\{\delta_i\}$ to identify the pairing with the shortest distance.  Only one qualifies.  Step 1 must find and delete the remaining $|\{\mu_{i+1}\}|-1$ possibilities.

Step 1 is not optimal even though it enumerates all possible mappings from FLSs in $\{\delta_i\}$ to the vacant coordinates in $\{\mu_{i+1}\}$.  This is trivial to demonstrate with an example.
Assume $\{\delta_i\}$=$\{P_1,P_2\}$ and $\{\mu_{i+1}\}$=$\{Q_1,Q_2\}$.
Assume distance from $P_1$ to each of $Q_1$ and $Q_2$ is 1 and 2 cells, respectively.  And, distance from $P_2$ to each of $Q_1$ and $Q_2$ is 2 and 5 cells, respectively.
The optimal minimum distance of 4 is realized by flying the FLS located at $P_1$ to the coordinates of $Q_2$, and flying the FLS at $P_2$ to the coordinates of $Q_1$.
However, Step 1 maps $P_1$ to $Q_1$ because their distance is the smallest.
It must subsequently map $P_2$ to $Q_2$, resulting in 6 as the total travelled distance by FLSs at $P_1$ and $P_2$.
This example demonstrates Step 1 of Simple does not compute optimal pairings of FLSs in $\{\delta_i\}$ and $\{\mu_{i+1}\}$.

\subsubsection{Motion Illuminations, Motill, Encoding}\label{sec:motill}
Motill is a family of divide-and-conquer encoding techniques to implement Step 1 of Simple.
%They detect and resolve Scenarios~\ref{simple:scenario1}-\ref{simple:flyoldnewcolor} between two point clouds $\Xi_i$ and $\Xi_{i+1}$.
%They assume an FLS rendering a point in $\Xi_i$ travels a short distance to render a different point in $\Xi_{i+1}$.
They localize mapping of the points by constructing a 3D grid on the point clouds.
Instead of computing the distance between a freed FLS of $\Xi_i$, $\{\delta_i\}$, with every vacant coordinate of $\Xi_{i+1}$, $\{\mu_{i+1}\}$, Motill compares those in the same cuboid or its neighboring cuboids.  Other pairings are guaranteed to be farther away.  By eliminating them from consideration, Motill reduces complexity of Step 1 to provide a faster execution time when computing FLS flight paths.
In addition, 
a Motill technique named ICF provides shorter flight distances for FLSs when compared with Simple. 

Motill is highly parallelizable and may use multiple cores of a processor to compute flight paths, see Section~\ref{sec:parallel}.
Simple is a special case of Motill with a grid consisting of 1 cuboid, see Section~\ref{sec:motillandsimple}.

Both lossy and lossless variants of Motill are possible.  In lossy mode, Motill may remove certain points to minimize the number of FLSs used to render an illumination.  In lossless mode, the number of FLSs is the same as the number of points.  The focus of this papers is on the lossless Motill. 

Motill partitions a scene consisting of $n$ points clouds into a Group of Point Clouds (GPCs).
Each GPC consisting of $\omega$ point clouds, $\{\Xi_1, \Xi_2, \cdots, \Xi_\omega\}$, $\omega \leq n$.  A point in a GPC corresponds to an FLS.  Motill computes flight paths for different FLSs across a GPC $\{\Xi_1, \Xi_2, \cdots, \Xi_\omega\}$.  Subsequently, Motill combines flight paths computed for the different GPCs together to compute the travel path of FLSs for the entire scene.

Motill processes a GPC by constructing a 3D grid on its first point cloud $\Xi_1$.  A maximum limit $\theta$ is imposed on the number of points assigned to each cuboid of the grid.  Every time a cuboid overflows, Motill breaks the cuboid into two by partitioning it along a dimension.  In our current implementation, we use a round-robin policy to select among the dimensions across all cuboids.  However, it is possible to develop more sophisticated policies to better balance points across the cuboids. 

Motill constructs a copy of the $\Xi_1$ grid with $\rho_1$ cuboids on the remaining point clouds of a GPC, $\Xi_2$ to $\Xi_\omega$.
%Motill superimposes a copy of the $\Xi_1$ grid with $\rho$ cuboids on the remaining point clouds of a GPC, $\Xi_2$ to $\Xi_\omega$. 
It scans points of $\Xi_i$ ($1 < i \leq \omega$) and assigns each to the cuboid that contains it, populating $\Xi_i$ grid.
%%It populates the grid copy of $\Xi_i$ ($i>1$) by scanning its points and assigning each point to the cuboid that contains it.
This step does not detect overflows and has no cuboid splits.
Hence, the cuboids of these point clouds may have more points than the maximum limit $\theta$.

The purpose of the grid is to reduce the number of points considered when computing the shortest distance.  Its cuboids localize how a point (FLS) changes position from one point cloud to the next.  By using the same grid across all point clouds of a GPC, Motill localizes changes to a few cuboids.
To describe how this is accomplished, 
%%%see Section~\ref{sec:gridprocessing}. 
we define the terms grid and neighboring cuboids.  We use these terms in the following section that details Motill's processing of the grid.
%%%describe how Motill uses the grid to detect Scenarios~\ref{simple:scenario1}-\ref{simple:flyoldnewcolor} and compute FLS flight paths to address them.

%\begin{definition}
%A cuboid consists of six flat faces and eight vertices.  All its faces are rectangles.  A cuboid is a square prism when at least two faces are squares.  A cuboid is a cube when all its six faces are squares.  
%A Motill grid may consist of a mix of cuboids, square prisms, and cubes.  We use the term cuboid to refer to an element of a grid even though it may be a cube or a square prism.  
%\end{definition}

\begin{definition}
A grid is a 3-dimensional coordinate space consisting of a fixed number of cuboids $\rho$.  
\end{definition}

\begin{definition}\label{def:neighbor}
Two cuboids of a grid are neighbors if their coordinate spans 
overlap along 2 dimensions (i.e., share a 2D face either partially or fully) and abut along one dimension.   
\end{definition}
%For example, in Figure~\ref{fig:intro:rosegrid}b, Cuboid 2 is neighbors with 1, 3, and 5.  Cuboid 3 has all the shown cuboids as its neighbor.  

\subsubsection{Grid processing}\label{sec:gridprocessing}
Motill processes the cuboids that constitute the grid of $\Xi_1$ to $\Xi_\omega$ to compute FLS flight paths, $\{\epsilon_1, \epsilon_2, \cdots, \epsilon_{\omega-1}\}$.  It processes two sequential point clouds, $\Xi_i$ and $\Xi_{i+1}$, by enumerating their respective cuboids and processing one pair at a time.
We identify one such pairing as $C_i^j$ and $C_{i+1}^j$.  The subscript identifies the point cloud and the superscript identifies the cuboid.  A pairing must have the same superscript value, i.e., identify the same cuboid in two consecutive point clouds with identical L, H, and D dimensions and neighbors.

% \begin{enumerate}
    % \item No change: $C_i^j$ and $C_{i+1}^j$ have the same number of points.  Both the L, H, D coordinate of each point and its color is the same across both cuboids.
    
    % \item Color change: $C_i^j$ and $C_{i+1}^j$ have the same number of points and the coordinates of their points is identical.  However, the color of some points is different.
    
    % \item Position change: While $C_i^j$ and $C_{i+1}^j$ have the same number of points and the color of these points is identical, the L, H, D coordinates of some points is different.
    
    % \item Different number of points:  The coordinates of the points and their color is the same for $C_i^j$ and $C_{i+1}^j$.  However, the number of points is different.  $C_i^j$ has either more or fewer points than $C_{i+1}^j$. 
    
    % \item Any combination of the previous three possibilities.  Six combinations are possible.  One has all three changes where $C_i^j$ and $C_{i+1}^j$ having different number of points, and position and color of some points change from $C_i^j$ to $C_{i+1}^j$.  
% \end{enumerate}

Motill uses Simple's Step 1 hashing technique to detect Scenarios~\ref{simple:scenario1}-\ref{simple:flyoldnewcolor}
for each cuboid pair $C_i^j$ and $C_{i+1}^j$.
With Scenario~\ref{simple:flyoldnewcolor},
it constructs $\{\delta_i^j\}$ and $\{\mu_{i+1}^j\}$ for each pairing
per discussions of Step 1, see Section~\ref{sec:simple}.

Motill computes intra-cuboid and inter-cuboid flight paths to transition FLSs of $\Xi_i$ to illuminate $\Xi_{i+1}$.
Intra-cuboid flight paths are local to cuboids $C_i^j$ and $C_{i+1}^j$.
Inter-cuboid flight paths may be in two from:  FLSs from a different cuboid in $\Xi_{i}$ flying into $C_{i+1}^j$ or FLSs from $C_i^j$ flying to a different cuboid in the point cloud $\Xi_{i+1}$.

%It computes intra-cuboid flight paths from a $C_i^j$ point to a $C_{i+1}^j$ point.  It computes inter-cuboid flight paths for FLSs that fly either from a different cuboid to $\Xi_{i+1}$'s $C_{i+1}^j$ or from $\Xi_i$'s $C_i^j$ to a different cuboid of $\Xi_{i+1}$.

Motill computes intra-cuboid flight paths by processing $\{\delta_i^j\}$ and $\{\mu_{i+1}^j\}$ using Simple, i.e., enumerating all possible combinations, sorting, and selecting the one with the shortest distance for its flight path.  
It computes inter-cuboid flight paths by identifying those cuboids of $\Xi_{i+1}$ that have more points than their respective $\Xi_i$ cuboids and those cuboids of $\Xi_{i+1}$ with fewer points than their respective $\Xi_i$ cuboids.  This results in two sets of cuboids, $\{C^+\}$ and $\{C^-\}$.  It processes one cuboid of $\{C^+\}$ by identifying its neighbors %(see Definition~\ref{def:neighbor}) 
to determine which appears in the set $\{C^-\}$.  For each such cuboid, it computes all possible mappings to select a destination point with the shortest distance.  

It is possible that there is no neighboring cuboid that intersects a cuboid in set $\{C^-\}$.  Motill collects these and in a final pass enumerates the points in cuboids of $\{C^+\}$ and points in cuboids of $\{C^-\}$.  It uses Simple's technique to map these points together, computing flight paths for FLSs.  

\subsubsection{Variants of Motill}\label{sec:variants}
Motill is a family of techniques.  In this paper, we consider two variants:  Intra-Cuboid-First (ICF) and Intra-Cuboid-Last (ICL).
ICF computes intra-cuboid flight paths for every cuboid pairing first.
For the remaining cuboids, it computes inter-cuboid flight paths.  It is motivated by the insight that computing flight paths local to a cuboid minimizes distance.

ICL reverses the order of these two steps, computing inter-cuboid flight paths first and intra-cuboid flight paths last.
Its motivation is that computing inter-cuboid flight paths first has the benefit of more candidate FLSs in $\{\delta_i^j\}$ and vacant destinations in $\{\mu_{i+1}^j\}$.
%ICF imposes a higher constraint on the inter-cuboid flights by removing candidates from $\delta_i^j$ and $\mu_{i+1}^j$.  

%With both variants, Motill has a final step for two consecutive points clouds, $\Xi_i$ and $\Xi_{i+1}$, with varying number of points.  Its computation identifies either cuboids of $\Xi_i$ with FLSs that are no longer needed in $\Xi_{i+1}$ or cuboids of $\Xi_{i+1}$ that require FLSs.  With the former, Motill may maintain the extra FLSs and require them to go dark, using them for the subsequent point cloud $\Xi_{i+2}$.  With the latter, Motill may either deploy FLSs from a dispatcher or use dark FLSs from $\Xi_{i+1}$.

Both ICF and ICL implement Step 2 of Simple.
%%% that considers the $\delta_i$ associated with every point cloud $\Xi_i$ to decide whether extra FLSs fly to a charging station or go dark for use by a future point cloud. 
Experimental results of Section~\ref{sec:evalmotill} show ICF is superior to ICL for the Rose illumination.
It executes faster and minimizes the total distance travelled by FLSs.  

\subsubsection{Parallelism}\label{sec:parallel}  
Motill variants may employ parallelism at different granularity, using multi-core CPUs with minimal coordination.
First, they may process different GPCs that constitute a scene in parallel.
A challenge is how to fuse travel paths computed for two consecutive GPCs, $\varphi_1$ and $\varphi_2$.
%There are several possible ways to fuse travel paths computed for two consecutive GPCs, $\varphi_1$ and $\varphi_2$.
%If each point of a point cloud is identified with a unique identifier across the point clouds then the task is trivial.  Motill simply takes the travel path computed for each unique point (FLS) of $\varphi_1$ and $\varphi_2$ in a sequence and merges them into one.  If no unique identifier exists then Motill may
A simple technique is to repeat the last point cloud $\Xi_\omega$ of $\varphi_1$ as the first point cloud $\Xi_1$ of $\varphi_2$.
%use the coordinates of the points in the last point cloud $\Xi_\omega$ of $\varphi_1$ to identify the corresponding points in the first point cloud $\Xi_1$ of $\varphi_2$.  
%The complexity of this matching task may be reduced by using the last point cloud $\Xi_\omega$ of $\varphi_1$ as the first point cloud $\Xi_1$ of $\varphi_2$.  
This reduces the matching task to a simple 3D coordinate lookup of the first point cloud $\Xi_1$ of $\varphi_2$ using the coordinates of the last point cloud $\Xi_\omega$ of $\varphi_1$.  For each match, the travel path for an FLS in $\varphi_1$ is concatenated with travel path from $\varphi_2$ after removing the redundant path attributed to the repeated use of $\Xi_\omega$ of $\varphi_1$.

Second, Motill may populate
the grids on $\Xi_2$ to $\Xi_\omega$ in parallel, using a different core for each point cloud.
(Recall from Section~\ref{sec:gridprocessing} that the structure of these grids is a copy of $\Xi_1$ grid.)  

Third, Motill may compute intra-cuboid flight paths for the different cuboids in parallel.  With hundreds of cuboids in a grid, Motill may use hundreds of cores concurrently.  Parallel processing of inter-cuboid flight paths requires an extra pre-processing step to identify those cuboids considered concurrently to not share neighbors.  %It is trivial for Motill to quantify the number of FLSs that may be removed or added to a cuboid by analyzing its number of points across the point clouds $\Xi_1$ and $\Xi_{i+1}$.

Motill may use the concept of stragglers to enhance latency.  Given a thousand core processor, assuming certain cores are idle, Motill may use these cores to execute a copy of tasks that are taking too long to complete (consuming the result of the copy that finishes first).
Stragglers waste computing resources by performing redundant work to enhance latency.

\subsubsection{Motill and Simple}\label{sec:motillandsimple}
%MOVE TO THE END OF THIS SECTION?
Motill with one cuboid emulates Simple.
We realize this by setting the capacity of a cuboid to a large number, i.e., maximum integer value.  This transforms the different variants of Motill to employ simple for intra-cuboid mappings.

Motill is faster than Simple.  
When configured with a reasonable number of cuboids for its grid, Motill also provides better flight paths that minimize total travelled distance.
We quantify this in the next section. %Section~\ref{sec:evalmotill}.

\subsubsection{A Comparison}\label{sec:evalmotill}
This section compares Simple, ICF, and ICL using the Rose illumination of Figure~\ref{fig:intro:rosegrid}a.
This comparison uses the same software, an implementation of Motill, for all three techniques.
An input parameter of Motill switches the order of intra-cuboid processing to be either first or last.  We emulate Simple by setting the capacity of Motill cuboids to the maximum integer value to force it to construct one cuboid.
We quantify both the execution time of an algorithm and the total distance of its computed flight paths.  The actual execution of the flight paths by FLSs to render an illumination and its latency is a part of our future work.

The reported execution times for Simple do not include the time to construct Motill's one cuboid grid.
With ICF and ICL, the reported execution times include the grid construction times.

All reported execution times were gathered from MATLAB R2022a (9.12.0.1884502) running on a MacBook Pro configured with a 2.3 GHz 8-Core Intel i9 processor and 16 GB of memory.
Its operating system is macOS Big Sur Version 11.5.2.
All the data was staged in memory prior to gathering execution times, eliminating disk I/O times.
While obtained results are written to disk, this is performed at the end after the experiment timing has stopped.
Reported experiments did not use parallelism.

\begin{figure}
\begin{subfigure}[t]{0.47\columnwidth}
\centering
\includegraphics[width=\textwidth]{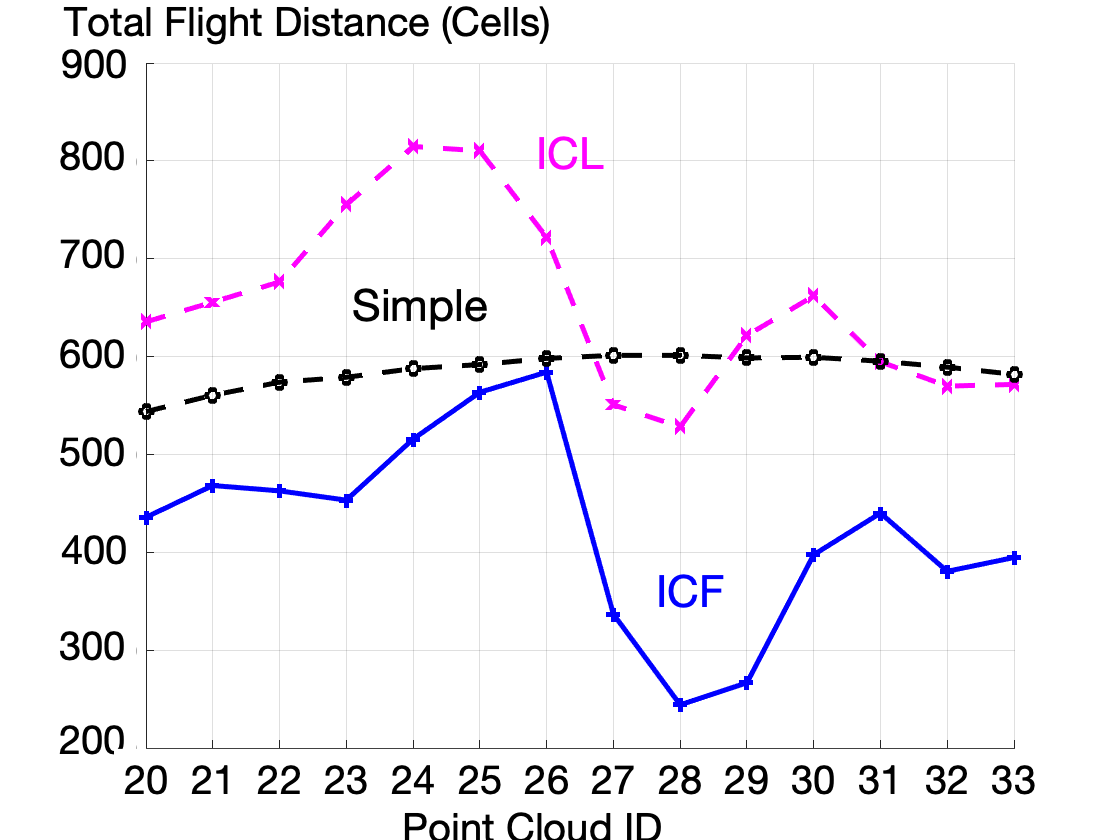}
\caption{Flight Distance.}
\label{fig:eval:ICLvsICFdist}
\end{subfigure}
\quad
\begin{subfigure}[t]{0.47\columnwidth}
\centering
\includegraphics[width=\textwidth]{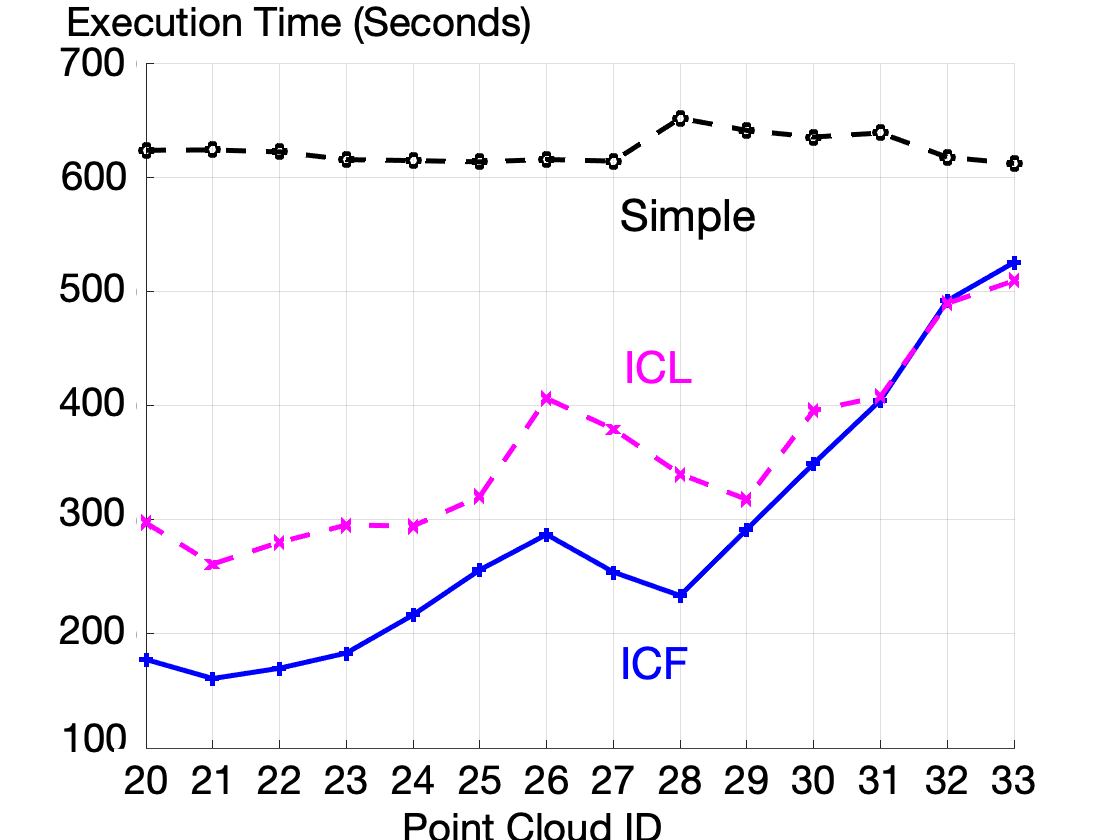}
\caption{Execution Time.}
\label{fig:eval:ICLvsICFtime}
\end{subfigure}
\caption{%%%Throughput as a function of $\beta$ with 
ICL vs. ICF, cuboid size ($\theta$) is 1500 points/FLSs.}
\label{fig:eval:ICLvsICF}
\end{figure}

\begin{figure}
\begin{subfigure}[t]{0.47\columnwidth}
\centering
\includegraphics[width=\textwidth]{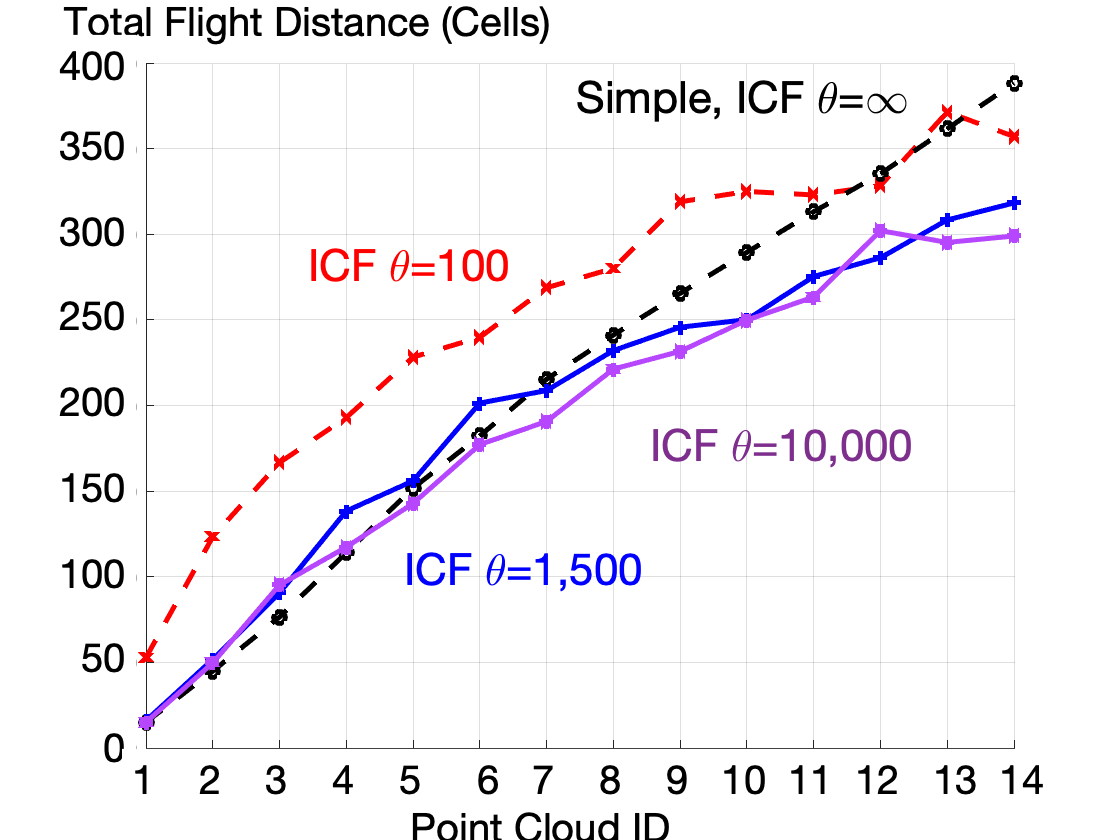}
\caption{ICF Flight Distance.}
\label{fig:eval:cubeSZdist}
\end{subfigure}
\quad
\begin{subfigure}[t]{0.47\columnwidth}
\centering
\includegraphics[width=\textwidth]{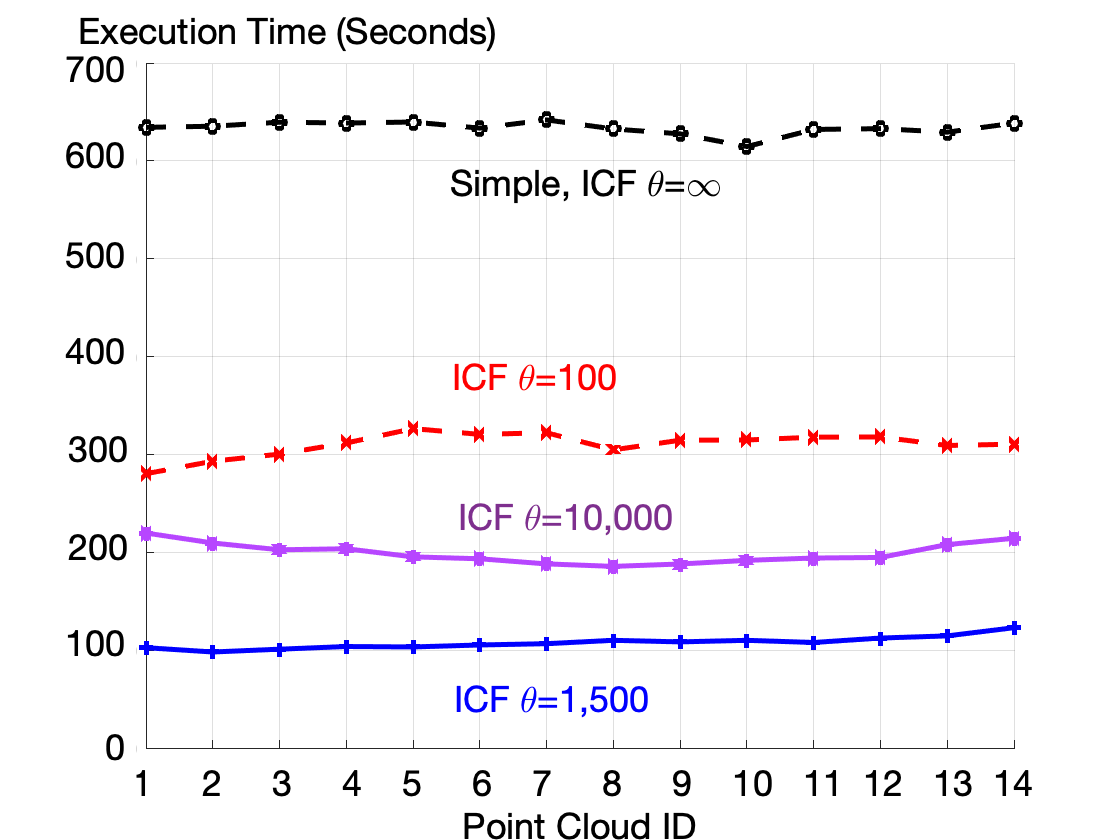}
\caption{ICF Execution Time.}
\label{fig:eval:cubeSZtime}
\end{subfigure}
\caption{%%%Throughput as a function of $\beta$ with 
ICF with different cuboid size ($\theta$) values.}
\label{fig:eval:cubesize}
\end{figure}

Main lessons are as follows:
\begin{itemize}
    \item ICF's computed flight paths provide a shorter total flight distance when compared with ICL.  Its execution time is faster for almost all point cloud pairings.  In those few cases that it is slower, the percentage difference is less than 4\%.  See discussions of Figure~\ref{fig:eval:ICLvsICFtime}.
    
    \item Simple is slower than both ICF and ICL.  With some point clouds, ICF is 5.5 to 6x faster than Simple.  See discussions of Figures~\ref{fig:eval:ICLvsICFtime} and~\ref{fig:eval:cubeSZtime}. 
    
    \item ICF provides comparable flight distances to Simple for most but not all pairs of point clouds.  See discussions of Figures~\ref{fig:eval:ICLvsICFdist} and~\ref{fig:eval:cubeSZdist}.
    
    \item Motill's cuboid size impacts its execution time and computed flight distances.  This is true with both ICF and ICL.  See discussions of Figure~\ref{fig:eval:cubesize}.
    
    \item With the Rose illumination, once FLSs are deployed using either MinDist or QuotaBalanced, the flight paths computed by the different techniques do not result in collisions.  Hence, no FLS collisions are reported.
    
    \item The point clouds that constitute the Rose illumination have the same number of points (65,321).  Hence, Simple and Motill do not execute their Step 2.
\end{itemize}

%\subsubsubsection{ICL vs. ICF}\label{sec:icl}
\noindent{\bf ICF vs. ICL:}
Figure~\ref{fig:eval:ICLvsICFdist} shows the flight distance computed by ICL and ICF.
The x-axis identifies the point cloud in the Rose illumination.  We used the 20$^{th}$ point cloud to the 33$^{rd}$ point cloud.
The y-axis shows the total distance of the flight paths computed by different techniques to transition from rendering point cloud $i$ (say 20) to the next point cloud $i+1$ (21).  All techniques compute the same number of flight paths (2009) for each point cloud.

Figure~\ref{fig:eval:ICLvsICF} highlights several observations.
First, ICF results in a lower flight distance for all point clouds.
Second, ICF is faster than ICL for most but not all the point clouds, see Figure~\ref{fig:eval:ICLvsICFtime}.
It is slightly (<4\%) slower than ICL for the 33rd point cloud.
Third, the point cloud data impacts FLS flight distances and Motill execution times significantly.
It is interesting to note that the flight distance of ICF increases linearly with the first 14 point clouds, see Figure~\ref{fig:eval:cubeSZdist}.
And, the total distance for each of these points is lower than that of point clouds 20-26 of Figure~\ref{fig:eval:ICLvsICF}.
The y-axis scale is different for Figures~\ref{fig:eval:ICLvsICFdist} and~\ref{fig:eval:cubeSZdist}.

%%%Both metrics vary for different point clouds.  To illustrate, compare the metrics reported for ICF in Figure~\ref{fig:eval:ICLvsICF} with those reported in Figure~\ref{fig:eval:cubesize}.  (Compare the blue lines in the presented graphs.)  They are different because Figure~\ref{fig:eval:cubesize} reports results for the point clouds 1 to 14. 

\noindent{\bf Cuboid size:}
The cuboid size impacts Motill's execution time and computed flight distances, see Figure~\ref{fig:eval:cubesize}.
It dictates the structure of the grid and the number of cuboids that constitute it.
A small cuboid size results in a large number of cuboids and slow execution times.
A large cuboid size is also undesirable because it results in a few cuboids and starts to approximate Simple.
It computes many combinations of possible flight paths only to select one.
 
Figure~\ref{fig:eval:cubesize} highlights this using ICF with the 1$^{st}$ to 14$^{th}$ point clouds of the Rose illumination.
We report on cuboid sizes 100, 1500, 10K, and max integer (1 cuboid).
(The latter is Simple.)
Cuboid size 1500 provides the fastest execution times, see Figure~\ref{fig:eval:cubeSZtime}.  Moreover, its computed flight distances are comparable if not better than Simple, see Figure~\ref{fig:eval:cubeSZdist}.

In Figure~\ref{fig:eval:cubesize}, 
$\theta=100$ results in 987 cuboids.
The portion of execution time used by ICF is 18\%.  The rest of the time is spent constructing and copying the grid.  It is interesting to note that the time to copy and populate the grid with point clouds 2 to 14 is approximately the same as constructing the grid on point cloud 1.

In contrast, $\theta$=1500 constructs 68 cuboids.  Its time to construct the grid on the first point cloud is twice the time to copy it on the remaining point clouds.  Both are approximately 6x faster than $\theta$=100. The portion of execution time used by ICF increases to approximately 75\%. 

A large cuboid size, $\theta$=10,000, constructs only 10 cuboids.
It spends 90\% of its time computing possible flight paths that it subsequently discarded.

Note that $\theta$=10,000 computes competitive flight paths comparable to those of $\theta$=1,500 and is superior to Simple, see Figure~\ref{fig:eval:cubeSZdist}.
It also provides execution times that are at least 3x faster than simple, see Figure~\ref{fig:eval:cubeSZtime}.

Simple is the extreme cuboid size, resulting in a grid with one cuboid.
Its execution time is the highest in Figure~\ref{fig:eval:cubeSZtime}.
It does not incur the overhead of constructing a grid.
Instead, 
it wastes time and resources by computing many FLS flight paths that are subsequently deleted.  See discussions of Section~\ref{sec:simpleanalysis}.

\section{FLS Failure Handling}\label{sec:failure}

An FLS is a mechanical device that may fail.  Its failure may degrade the quality of a rendering by not illuminating one or more of its points.
There are several types of FLS failures:
rotor failures,
light source failures, 
computing failures in the form of reboots,
and battery power failures.
Assuming these failures are independent and occur at a constant rate, one may compute the Mean Time To Failure of an FLS ($MTTF$) similar to how magnetic disk manufacturers calculate the MTTF of disk drives~\cite{gibson}.
The Mean Time to Degraded Illumination (MTDI) is a linear function of the number of FLSs ($\alpha$) that constitute an illumination:  $MTDI=\frac{MTTF~of~1~FLS}{\alpha}$.

%A failed FLS results in loss of its assigned lightings and flight paths, degrading the quality of an illumination.  The Mean Time to Degraded Illumination (MTDI) is a linear function of the number of FLSs ($\alpha$) that constitute an illumination:  $MTDI=\frac{MTTF~of~1~FLS}{\alpha}$. 

Assuming an FLS fails once a month (MTTF of 720 hours), the MTDI of the Rose illumination with $\alpha$=65,321 FLSs is 40 seconds.
This is disheartening if we want to scale to illuminations consisting of millions of FLSs.
Below, we describe a group parity/replication technique to enhance MTDI of an illumination in the presence of frequent FLS failures.  We start by describing how FLSs cooperate to detect failures.  Subsequently, we describe grouping of FLSs with a standby using data replication and parity techniques to improve MTDI.  We discuss group formation techniques and analyze MTDI as a function of group size.

\noindent{\em Failure detection:}  FLSs cooperate to detect failures and notify the Orchestrator of the identity of the failed FLS.
This cooperation is in two forms.
First, once an FLS detects its own failure, it uses its networking to inform its neighbors and the Hub (Orchestrator) of its failures.
%First, once an FLS detects its own failure, it communicates this information with its neighbors and the Hub (Orchestrator) using its networking capability. 
This applies to the first two forms of failures.  
With light source failures, the FLS flies to a Terminus immediately as it is no longer able to illuminate a point.
With rotor failures, it repels~\cite{roguedrone2021,repel1,repel2} FLSs in its downward descent by generating frequent failed messages.  Those FLSs that receive this message move away to prevent the failed FLS from crashing into them.  With the architecture of Figure~\ref{fig:arch}, the failed FLS falls on the conveyor belt of the garbage collector and is deposited into a Terminus.  

Second, FLSs exchange periodic heartbeat messages~\cite{can2001,chord2001} with their neighbors in the display mesh to detect processor and battery failures.
An FLS that encounters these forms of failures may not be able to notify other FLSs of its failure.
Hence, FLSs cooperate to detect these failed FLSs.  
%This is because an FLS that encounters such failures 
%The last two forms of failures prevent an FLS from notifying other FLSs.  
%The detection mechanism requires FLSs to exchange periodic heartbeat messages with their neighbors in the display mesh.  
If an FLS does not receive an anticipated heart beat message from one of its neighbors then it polls the neighbor.
After a few failed attempts, it identifies the neighbor as having failed and communicates the identity of this failed FLS to its neighbors and the Hub (Orchestrator).
%This information propagates throughout the FLS swarm and the display Hub to the Orchestrator.  

\noindent{\em Parity and replication groups:}  To maintain the quality of an illumination, a failed FLS must be replaced with a new one quickly.
The system must restore both the physical FLS and the data that describes its flight path and lighting responsibilities.
We assume this data is stored on the local storage of each FLS.

We use standby FLSs to recover from FLS failures.
In normal mode, standby FLSs are dark.  After the discovery of a failed FLS, the standby assumes the lighting responsibilities and flight paths of the failed FLS.
This occurs concurrently with the Orchestrator deploying a replacement FLS to substitute for the standby.

We use parity and replication techniques to maintain the data 
%pertaining to the lighting responsibilities and flight paths (i.e., data) 
of a failed FLS available.
These techniques assign $G$ FLSs in close proximity of one another to a group and assigns one or more standbys to each group.
The minimum value of $G$ is one, requiring
a standby for every FLS that is illuminating a point.
This standby has a copy of the data assigned to its paired FLS and mirrors its flight paths.
If an FLS fails, its standby resumes its lighting and flight responsibilities immediately using its local copy of the data.  
A challenge of $G$=1 is how to prevent the standbys from obstructing the user's field of view (FoV).

With $G>1$, replication requires the standby FLS to maintain a copy of the data assigned to each FLS in its group.
A parity technique requires the standby to compute the parity (xor) of the data assigned to the $G$ FLSs in the group.
% representing flight paths assigned to the $G$ FLSs.
%The resulting parity data is the same size as the largest flight data assigned to an FLS in the group.

Parity techniques are more space efficient than replication techniques.
However, a parity technique requires the standby to fetch data from the remaining $G-1$ FLSs and use this data in combination with its parity data to compute the missing data of a failed FLS.
%%However, to compute the missing data of a failed FLS in the group, they require the standby to obtain and use the flight paths of the remaining $G-1$ FLSs.
%%%Using this data in combination with its parity data, the standby computes the flight paths and illumination responsibilities of the failed FLS.  
With more than 1 failure in a group, the standby must wait for the Orchestrator dispatched FLSs to arrive with their data before it is able to compute the missing data.
%A large value of $G$ increases the probability of a second failure before the group returns to normal mode of operation.
With replication, the standby may select a failed FLS and use the copy of its data to substitute for the failed FLS.  
%The illumination quality may suffer until the Orchestrator's dispatched replacements arrive and substitute for the other failed FLSs. 
%With a parity technique, if the standby computes the missing data prior to the second failure then it may proceed to substitute for the failed FLS.  Otherwise, it waits until the Orchestrator's dispatched replacement with the missing data arrives to enable it to compute the missing data.

%%%\subsection{Group Construction}
\noindent{\em Group construction:}
The $G$ FLSs in a group should be in close proximity of one another.  This minimizes the distance travelled (time) by a standby to substitute for a failed FLS in the group.  If the group is using a parity scheme, close proximity facilitates local communication 
%between the standby and the remaining $G-1$ FLSs 
for the standby to obtain data from the remaining FLSs
%their flight paths and lighting responsibilities 
to compute the failed FLS's flight path and lighting responsibilities. 

FLS group construction is a weighted matching problem~\cite{tutte47,edmonds65}.
There exists centralized~\cite{avis83,preis99}, distributed~\cite{hoepman2004}, and decentralized~\cite{kclique2014} algorithms for this problem.
With static illuminations, one may adapt these for use by the Orchestrator, multiple dispatchers, and millions of FLSs, respectively.
This is a short-term future research direction.

With motion illuminations, the position of some FLSs will change from one point cloud to the next.  This may change the distance between FLSs that constitute a group.
The FLSs may re-construct the groups using decentralized~\cite{kclique2014} algorithm.
This may require re-assignment of standbys.  With a parity technique, an impacted standby must re-compute its parity information for the new group.
With replication, the standby must delete flight paths and lighting pattern of FLSs that it is no longer responsible for and obtain a copy of the flight paths and lighting patterns of the new FLSs that it may substitute for.
To minimize the amount of exchanged data, a display may relax the value of $G$ for evolving groups, allowing some groups to consist of more than $G$ FLSs and others to consist of fewer than $G$ FLSs.   

Another approach is to maintain the same FLS membership for the group and require the standby to adjust its position to approximate the same distance to the different FLSs that constitute the group.  If the FLS reliability is heterogeneous, the standby may position itself closer to those FLSs with a higher failure probability.  

Yet another possibility is to remove the constraint that each FLS must belong to a group.  This enables an FLS to leave a group, provide its illumination for a few point clouds without being a part of a group, and possibly join another group.  Once it joins a group, the standby of the impacted groups must adjust its membership information and replicated/parity data.  An FLS crucial to the illumination may be provided with a mirror standby.
An investigation of these possibilities is a future research direction.

%\subsection{MTDI using Groups}
\noindent{\em MTDI with different group sizes:}
We adapt the models of~\cite{gibson} to establish the MTDI of an illumination as a function of group size using a parity technique.
MTTF of a group is
$MTTF_{Group}$=$\frac{MTTF_{FLS}}{G+1}\times \frac{1}{P}$,
where
$P$ is the probability of another FLS failure in a group before restoring the group to normal mode of operation,
$P=\frac{MTTR}{MTTF_{FLS}/(G+1)}$.
MTTR is the Mean Time To Repair a group with a failed FLS to normal mode of operation.  It is the amount of time elapsed from when an FLS fails to the time the group is restored to normal mode of operation with a replacement FLS.
The Mean Time to Degraded Illumination is:
$MTDI=\frac{G \times MTTF_{Group}}{\alpha}$ where $\alpha$ is the number of FLSs to render an illumination.

Table~\ref{tbl:parity} shows MTDI of the Rose illumination assuming an FLS fails once a month and the system's MTTR is 1 second.
Reliability groups enhance MTDI from 40 seconds to almost two months with 20 FLSs in a group, $G$=20.
This is enhanced almost two folds (to 111 days) with 10 FLSs per group, $G$=10.
With $G$=10, the Rose illumination requires approximately 6,400 additional FLSs. 

\begin{table}
%\begin{small}
  \caption{Reliability Groups enhance MTDI of the Rose illumination from 40 seconds to more than a month.}
  \label{tbl:parity}
  \begin{tabular}{c|cc}
    \toprule
    & G=10 & G=20\\
    \midrule
    Total Number of FLSs & 71,853 (1.1$\alpha$) & 68,588 (1.05$\alpha$) \\
    Overhead Cost & 10\% & 5\% \\
    MTDI Hours (Days) & 2670 (111) & 1399 (58) \\
    %$\Psi^2_1$ & 1 in 40,000& Unexplained usage\\
  \bottomrule
\end{tabular}
%\end{small}
\end{table}

\section{Staggered Battery Charging, STAG}\label{sec:battery}

STAG is a novel algorithm that staggers charging of FLS batteries as a function of time 
%%%with the objective 
to minimize both the number of charging stations and the overall number of FLSs required to render an illumination.

STAG makes several assumptions.  First, each FLS has a battery that provides a finite flight time $\beta$ when fully charged.  Second, $\Omega$ time units are required to fully charge a depleted battery with minimal or no remaining flight time left.  Third, the time to charge an FLS battery, $\lambda$, is a linear function of $\beta$ and its remaining flight time $r$, $\lambda$=$\frac{\Omega*r}{\beta}$.  
%Fourth, STAG assumes a parity based technique using standby FLSs is used to tolerate FLS failures.  The battery of these standbys is similar to the other FLSs.  

An FLS computes the amount of battery flight time required for it to fly to a charging station using its distance to the charging station.  Once its battery flight time reaches this threshold, the FLS will go dark and fly to the charging station.

A standby FLS (see Section~\ref{sec:failure}) will substitute for this FLS to perform its lighting responsibility while a dispatcher will deploy an FLS with $\beta$ flight time to substitute for the standby.  %%The FLS flying back may be the standby for the group.  

It is important to minimize the window of time $\Delta$ for a battery depleted FLS$_d$ to switch places with a fully charged FLS$_c$.
Should an FLS belonging to the parity group of FLS$_d$ fail during $\Delta$, this may result in loss of information and a degraded illumination.  %We address this in Section~\ref{sec:STAGdiscuss}

{\em Preliminaries:}  Consider an illumination with $\alpha$ FLSs and assume the time to charge the battery of each FLS equals its flight time on a fully charged battery, $\Omega=\beta$.
A naive algorithm may deploy all $\alpha$ FLSs at one instance in time.  After $\beta$ time units, all FLSs must fly back to a charging station while the dispatchers deploy $\alpha$ fully charged FLSs.  Naive repeats this process every $\beta$ time units while rendering the illumination. 
  
Naive has several limitations.  First, it requires 2$\alpha$ FLSs to render an illumination:  $\alpha$ FLSs charge while $\alpha$ FLSs render the illumination.  Second, there is an exchange step when fully changed FLSs and fully depleted FLSs switch places.  The illumination may become distorted during this period because almost all FLSs that constitute an illumination may go dark in order to switch places with fully charged FLSs.  Moreover, the algorithm that manages FLSs during this period has a high complexity as it must manage flight patterns of 2$\alpha$ FLSs. 

{\em \bf STAG} staggers FLSs as a function of time to prevent them from exhausting their finite flight time at the same time.  It overlaps charging of some FLSs with others that are rendering an illumination, eliminating Naive's exchange step.  In its steady state, STAG switches a fully charged FLS with a fully depleted FLS continuously.  Details of STAG are as follows.

STAG constructs $h$ flocks of FLSs.  A flock $i$ consists of $\alpha_i$ FLSs.  Within a flock $i$, STAG staggers its $\alpha_i$ FLSs such that their remaining battery flight time ranges from $\beta$ down to the staggering interval S, S=$ \frac{\beta}{\alpha_i}$.  Assuming the FLSs in a flock are numbered from 1 to $\alpha_i$, the remaining flight time of FLS $j$ is $\beta(j)=\frac{j*\beta}{\alpha_i}$.  Thus, FLS $j=\alpha_i$ has a fully charged battery with $\beta$ flight time.

The number of FLSs charging (or staged in a hangar) to substitute for an FLS of Flock $i$ is $\lceil \frac{\Omega}{S} \rceil$.  This is the extra number of FLSs required by a flock to render an illumination.  The total number of FLSs that are charging (or staged in a hangar) is $h \times \lceil \frac{\Omega}{S} \rceil$.  %There is a startup overhead associated with STAG that we describe at the end of this section. 

\begin{theorem}
STAG minimizes the number of additional FLSs required to render an illumination.
\end{theorem}

\begin{proof}
An FLS spends $\beta$ time units rendering an illumination and $\Omega$ time units charging.  The fraction of time an FLS spends illuminating is $\frac{\beta}{\Omega+\beta}$.  The minimum number of FLSs, $\Psi$, to render an illumination consisting of $\alpha$ FLSs must satisfy the following equality:  $\frac{\beta}{\Omega+\beta} \Psi = \alpha$.
%\begin{displaymath}
%\frac{\beta}{\Omega+\beta} \Psi = \alpha 
%\end{displaymath}
Solving for $\Psi$, one obtains:
\begin{equation}\label{eq:addFLS}
\Psi=\alpha+\alpha\frac{\Omega}{\beta}
\end{equation}
Thus, a minimum of $\alpha\frac{\Omega}{\beta}$ additional FLSs are required to render the illumination.  

STAG requires $\frac{\Omega}{S}$ additional FLSs.  Substituting the definition of S, $\frac{\beta}{\alpha}$, in this equation produces $\alpha\frac{\Omega}{\beta}$.  This equals the minimum number of additional FLSs in Equation~\ref{eq:addFLS}.
\end{proof}

\begin{exmp}
\label{exp:eg1}
Consider an illumination consisting of 5 FLSs, $\alpha$=5.  These FLSs are assigned to 1 flock, $h$=1.  The time to charge an empty FLS battery is 3 minutes and each FLS provides 15 minutes of flight time on a fully charged battery, $\Omega$=3 and $\beta=15$.  STAG interleaves the charging of the 5 FLSs such that each FLS is staggered S=3 minutes of flight time apart, S=$\frac{\beta=15}{\alpha=5}$.  The extra number of FLSs is 2, $\lceil \frac{\Omega=5}{S=3} \rceil$.
While 2 FLSs are charging, 5 FLSs render the illumination.  Every 3 minutes, the Orchestrator deploys a fully charged FLS to render the illumination.  It replaces an FLS with an almost depleted battery that returns to a charging station to be charged.  This FLS spends 3 minutes in a charging station.  Subsequently, the Orchestrator deploys it to substitute for another FLS in the flock with an almost depleted battery.$\blacksquare$
\end{exmp}

The number of FLSs in transit from a charging station to an illumination is $2h$.
In Example~\ref{exp:eg1} with $h=1$, there are 2 FLSs in transit.  One flying from the illumination to a charging station and a second from a charging station to the illumination.
A display must ensure these FLSs are not in the user's field of view.

Given a swarm, reducing its number of flocks to one, $h$=1, minimizes the number of FLSs in transit.  This reduces the likelihood of a dark FLS from obstructing the user's field of view.  

However, the number of flocks is dictated by the maximum time required for an FLS to fly back to the charging station, $S_{threshold}$, and the number of FLSs required by an illumination $\alpha$.  
%%$S_{threshold}$ is the maximum flight time required by an FLS to travel to a charging station.
With an illumination that consists of a large number of FLSs, maintaining $h=1$ results in a small staggering interval S (because S is a function of the number of FLSs in a flock).  
$S_{threshold}$ dictates the number of FLSs in a flock ($\alpha_i=\frac{\beta}{S_{threshold}}$) which in turns dictates the number of flocks, $h=\lceil \frac{\alpha}{\alpha_i} \rceil$.  

Flocks that constitute an illumination may use different staggering intervals $S$ and consists of a different number of FLSs $\alpha_i$.
This is highlighted by the analysis of Section~\ref{sec:staganalysis}.

\subsection{Flock Formation} With STAG, flocks are logical.  Unlike the reliability groups of Section~\ref{sec:failure}, there is no constraint on them to be in close proximity of one another.
Each FLS requires a flock id $i$ and a logical number (termed {\em stag-id}) within that flock ranging from 1 to $\alpha_i$, 1 $\leq$stag-id$\leq \alpha_i$.
%However, similar to forming groups, this matching problem has well known centralized~\cite{avis83,preis99}, distributed~\cite{hoepman2004}, and decentralized~\cite{kclique2014} algorithms.  An FLS display may adapt one of these techniques to form STAG's flocks.%FLSs may implement and execute such a protocol to form flocks and implement STAG.

%A centralized technique %%for formation of flocks may require 
The Orchastrator may compute the number of flocks and the quota $\alpha_i$ for each flock.  It assigns flock ids to different dispatchers that deploy FLSs.  As a dispatcher deploys an FLS, it assigns 
a flock id and stag-id (a simple counter) to each FLS until the quota of the flock is exhausted, i.e., stag-id equals $\alpha_i$ of flock $i$. 
There will most likely be scenarios where multiple dispatchers are required to deploy FLSs of a single flock.  In this case, the Orchestrator assigns a unique range of stag-id values to each participating dispatcher.

\subsection{Staggering FLSs}

A display that initiates an illumination for the first time may have its FLSs with flight times that are not staggered.  How does the display stagger FLS flight times S time units apart given some initial flight time for its FLSs?  This section answers this question assuming initial flight time of $\beta$, i.e., all FLSs have fully charged batteries.  One approach is to deploy FLSs of a flock every $S$ time units.  A limitation of this approach is that it introduces a delay of $(\alpha_i-1) \times S$ to render an illumination.  The delay is 12 minutes with the small scenario of Example~\ref{exp:eg1}.  This delay may become unacceptable with large values of $\alpha_i$ and $S$.  With the 65,321 FLSs required by the Rose illumination, assuming $S$ is 1 second, the delay is more than 18 hours.

We now present a decentralized and efficient technique to stagger FLSs.  This technique deploys $\alpha_i$ FLSs that constitute a flock as fast as possible to render an illumination quickly.  Each $\alpha_i$ FLS is notified that this is the first time it is being deployed.  This causes FLS $j$ to fly back to a charging station after rendering the illumination for $j \times S$ time units, $j$ is the stag-id assigned to an FLS in a flock, $1 \leq  j \leq \alpha_i$.  FLS $j$ is replaced by another FLS with a fully charged battery that is no longer identified as being deployed for the first time.  This FLS performs flight and lighting responsibilities of FLS $j$ for the entire $\beta$ time units per earlier description of STAG.  

In essence, this technique requires $\alpha_i-1$ FLSs deployed for the very first time to fly back to a charging station with partially full batteries to charge fully.
%%%In essence, STAG requires $\alpha_i-1$ FLSs with partially full batteries to fly back to a charging station to become fully charged for the very first time they are deployed to render an illumination.  
By controlling the remaining battery life time of these FLSs prior to their flight back, this technique staggers flight time of %$\alpha_i$ 
FLSs in a flock to be $S$ time units apart.

In Example~\ref{exp:eg1}, STAG deploys all $\alpha_i$=5 FLSs as fast as possible.  For their first deployment, FLSs 1, 2, 3, 4, and 5 fly back after rendering the illumination for 3, 6, 9, 12, and 15 time units, respectively.  This is because the staggering interval is 3 time units, $S=3$.  STAG requires 7 FLS for this illumination, 2 charge while 5 render the illumination.  The 2 extra FLSs are in a hangar because they are fully charged.  FLSs 1 and 2 are replaced by these two FLSs.  FLSs 3 and 4 are replaced by FLSs 1 and 2 with fully\footnote{We assume the time to charge a battery is a linear function of its remaining flight time.  FLS 1 is fully charged after 3 time units while FLS 2 is fully charged after 6 time units.  This overlaps with the time FLSs 3 and 4 are rendering the illumination, enabling batteries of FLSs 1 and 2 to become fully charged.} charged batteries.  For the remainder of the illumination time, FLSs are staggered $S=3$ time units apart.

\begin{table}
\begin{small}
  \caption{STAG with the Rose illumination.% and different $\beta$ and $\Omega$ values.
  }
  \label{tbl:stag}
  \begin{tabular}{l|ccc}
    \toprule
    & $\beta$=5 min & $\beta$=10 min & $\beta$=20 min \\
     & $\Omega$=10 min & $\Omega$=5 min & $\Omega$=2.5 min \\
    \midrule
    $h$, number of flocks & 218 & 109 & 55 \\
    $\alpha_i$, FLSs/Flock & 300	& 600 & 1200 \\
    Extra FLSs per Flock & 600 & 300 & 150 \\
    \midrule
    FLSs for the last Flock & 221 & 521 & 521 \\
    S for the last flock (Millisec) & 1358 & 1152 & 2303 \\
    Extra FLSs for the last flock &	442 & 261 & 65.125 \\
    \midrule
    Extra FLSs for the illumination & 130642 & 32661	& 8166 \\
    Overhead Cost & 200\% & 50\% & 12.5\% \\
    Total Number of FLSs & 195,963 & 97,982 & 73,487 \\
    %$\Psi^2_1$ & 1 in 40,000& Unexplained usage\\
  \bottomrule
\end{tabular}
\end{small}
\end{table}

\subsection{An Analysis}\label{sec:staganalysis}
Table~\ref{tbl:stag} quantifies the behavior of STAG with different $\beta$ and $\Omega$ values.  The first column pertains to the flight time and battery charge time of today's Sky Viper Dash Nano Drone\footnote{Approximate cost of \$17, length=2 inches, width=2.5 inches, height=0.75 inches, weigth=7 ounces.}.  
The other two columns correspond to future generations of such a device with flight time on a fully charged battery ($\beta$) doubling and the time to charge ($\Omega$) its battery is halved. 

We set the lower bound on $S$ to 1 second.
It limits the number of FLSs in a flock, $\alpha_i$, shown in the 2nd row of Table~\ref{tbl:stag}.  The number of FLSs that constitute the Rose illumination ($\alpha$=65,321) is not an even multiple of $\alpha_i$.  Hence, the middle 3 rows show the characteristics of the last flock that has the remaining FLSs.  Note that its value of $S$ is higher than 1 second (1000 millisec) because it has fewer FLSs.  

The characteristics of today's Sky Viper battery increases the number of FLSs to render the Rose illumination 3 folds.  This is a 200\% overhead.  This overhead decreases linearly as we enhance battery characteristics,
see the 2nd to last row of Table~\ref{tbl:stag}.
With an 8 fold overall improvement in battery characteristics (4 fold enhancement of $\beta$ and 4 fold reduction of $\Omega$), STAG's overhead decreases to 12.5\%.

\section{Related Work}\label{sec:related}

FLS displays render virtual objects in a physical volume.  They are in the same class of systems as 
physical artifacts~\cite{sutherland1965} using programmable matter consisting of catoms~\cite{matter2005}, roboxels as cellular robots that dynamically configure themselves into the desired shape and size~\cite{roboxel1993}, fast 3D printing~\cite{t1000}, and BitDrones as interactive nano-quadcopter~\cite{gomes2016bitdrones}.
These studies describe 3D displays.  However, they do not present algorithms to render 3D illuminations.

FLS displays are inspired by today's indoor and outdoor drone shows that use illuminated, synchronized, and choreographed groups of drones arranged into various aerial formations.
An FLS display is similar because each FLS is a drone and a motion illumination is rendered by computing FLS flight paths that synchronize FLSs as a function of time and space.

Assignment of points to dispatchers (Section~\ref{sec:static}) and Step 1 of Simple that matches $\{\delta_i\}$ points to $\{\mu_{i+1}\}$ points (Section~\ref{sec:simple}) are centralized heuristics for euclidean matching~\cite{avis83}.
Both strive to solve a weighted matching problem~\cite{tutte47,edmonds65}.
There exists centralized~\cite{avis83,preis99}, distributed~\cite{hoepman2004}, and decentralized~\cite{kclique2014} algorithms for this problem.
MinDist is centralized and similar to the greedy heuristics of~\cite{kurtzberg62,tarjan81}.  QuotaBalanced is a novel greedy heuristic that balances the load of deploying FLSs across dispatchers.

A centralized, offline algorithm to compute a flight/lighting plan for outdoor light show performances is presented in~\cite{Sun2020PathPF}.  This algorithm requires drones to be placed in a field and in a specific arrangement.  It computes collision free paths for 500 drones to display different images in sequence, e.g., a ballerina followed by a guitar.  It may be modified for use by the Orchestrator to compute flight paths for point clouds that transition one scene to the next.  Motill is different because it is designed for a scene where changes from one point cloud to the next is not anticipated to be drastic.

There are many path planning algorithms for robots and UAVs~\cite{collisionfree2012,collisionfree2015,collisionavoidance2018,ReactiveCollisionAvoidance2008,ReactiveCollisionAvoidance2011,ReactiveCollisionAvoidance20112,reactiveColAvoidance2013,downwash1,downwash3,dcad2019,Engelhardt2016FlatnessbasedCF,navigation2017,reactiveColAvMorgan,reactiveColAvBaca,speedAdjust2021,gameCollisionAvoidance2020,gameCollisionAvoidance2017,preiss2017,Ferrera2018Decentralized3C,planning2019,preiss2017whitewash,opticalpositioning1,khatib1985,repel1,roguedrone2021}.
They address the challenge of moving from a given initial position to a set of predefined targets while avoiding collisions with obstacles as well as other UAVs.
Most relevant are studies that avoid UAV (robot) collision using
%The concept of cooperating drones (and robots) that avoid collisions is described in~\cite{khatib1985,repel1,roguedrone2021}. These studies avoid UAV (robot) collisions using 
an artificial potential field (APF) that defines a safety radius around the drone~\cite{khatib1985,repel1,roguedrone2021}.
With APF, the UAV (robot) moves to its target point guided by attractive force and repulsive forces.
These techniques are applicable to FLSs that execute the flight paths computed by Motill to render a motion illumination.  Our techniques have the added advantage that an FLS is informed of a potential conflict and FLSs may communicate to avoid a collision.
In~\cite{roguedrone2021}, a modified version of APF is used to identify a failing (rogue) drone.  
An FLS display may use this technique to detect a failed FLS.  

FLSs are network enabled and the failure detection techniques described in Section~\ref{sec:failure} are inspired by those used in peer-to-peer networks, e.g., CAN~\cite{can2001}, Chord~\cite{chord2001}.
Formation of FLS groups and use of standbys to tolerate failures is similar to disk striping techniques~\cite{striping86,gibson} in-use by the disk manufacturers.  
An interesting dimension introduced by FLSs is their flight paths that may change their memberships in groups.

\section{Future Work}\label{sec:future}
We are extending FLS displays in several exciting directions. 
First, we are developing a positioning system for the FLSs to compute their location in a display.
Such a positioning system is essential to enable FLSs to execute the flight paths computed by QuotaBalanced and Motill to render an illumination.  

Second, we are evaluating offline algorithms to form reliability groups using the computed FLS flight paths.
The offline algorithms compute membership of the mobile FLSs in a reliability group and provide this information for use by FLSs when rendering the illumination.
Their tradeoffs may involve metrics such as cost, traveled distance, consumed energy, amount of exchanged data, and MTDI.

Third, we are investigating the use of physics engines, e.g., Gazebo~\cite{Aguero-2015-VRC,Koenig-2004-394} and AirSim~\cite{airsim2017fsr} among others, to simulate the flight paths computed by ICF.
These simulation studies will implement the alternative collision avoidance techniques and quantify their tradeoffs.
They are the first step towards an implementation of an FLS display and its architecture.

Fourth, with Motill, STAG and our reliability techniques, it is important to keep dark FLSs out of the user's field of view.  An interesting question is the impact of the dark FLSs on the quality of an illumination perceived by a user.  This requires human studies using a prototype for different applications.  %An answer may be application dependent.

Fifth, a display may consist of heterogeneous FLSs with varying reliability, flight times on a fully charged battery, and time to fully charge their battery.  Both the reliability models and STAG are designed for FLSs with homogeneous characteristics.  These algorithms must be extended to support heterogeneous FLSs. %Novel algorithms and designs are required to support heterogeneous FLSs.

Finally, a swarm of FLSs may implement encounter-type haptic interactions~\cite{rodrigo2021} by generating force back against a user touch~\cite{abdullah2018hapticdrone,hoppe2018vrhapticdrones,auda2021flyables,abtahi2019beyond,shahram2022b}.
This will enable a user to see virtual objects as illuminations without wearing glasses and to touch them without wearing gloves~\cite{shahram2021}.
These concepts will facilitate immersive and interactive 3D displays depicted in science fiction shows, Star Trek’s holodeck.
We are developing algorithms for a swarm of FLSs to detect the location of user touch, quantify the amount of exerted force, and generate the kinesthetic sensation (“muscle sense” of contacting objects with mass) for a user.
User safety and trust are paramount and we intend to address them from the start~\cite{shahram2022b}.

\section{Acknowledgments}
We thank Hamed Alimohammadzadeh for generating the Rose illumination and his drawings.  We are grateful to the anonymous reviewers of the ACM Multimedia 2022 for their valuable comments. 

\bibliographystyle{ACM-Reference-Format}
\bibliography{main}  % vldb_sample.bib is the name of the Bibliography in this case

\end{document}